\documentclass[tbtags,reqno]{amsart}

\usepackage{color,caption}
\usepackage[noadjust]{cite}   
\usepackage{latexsym}
\usepackage{mathrsfs,amsmath,amsthm,amsopn,amsfonts,amssymb,epsfig,stmaryrd,mathtools,dsfont}
\usepackage{graphicx}
\usepackage{pstricks,tabmac}
\usepackage[latin1]{inputenc}
\usepackage[english]{babel}
\usepackage[T1]{fontenc}
\usepackage{enumitem}
	\setenumerate{label=\textup{(\arabic*)}} 
\usepackage[colorlinks=true,
            linkcolor=blue,
            urlcolor=blue,
            citecolor=black]{hyperref}
\newtheorem{example}{Example}[section]

\newtheorem{proposition}[example]{Proposition}

\newtheorem{conjecture}[example]{Conjecture}

\numberwithin{equation}{section}

\newcommand{\exterior}{\textstyle\bigwedge}

\def\be{\begin{equation}}
\def\ee{\end{equation}}

\def\lrw{\leftrightarrow}

\def\svir{\mathfrak{svir}}

\def\ta{\theta}
\def\La{\Lambda}
\def\Om{\Omega}
\def\Ga{\Gamma}
\def\Gab{\bar\Gamma}
\def\la{\lambda}
\def\a{\alpha}
\let\R\rangle\let\L\langle
\def\ct{\text{ct}}
\def\alpm{\alpha_{\pm}}
\def\ta{\theta}
\def\cd{\circledast}
\let\l\left\let\r\right
\let\d\partial

\newcommand{\tcercle}[1]{\ensuremath{\setlength{\unitlength}{1ex}\begin{picture}(2.8,2.8)\put(1.4,1.4){\circle{2.8}\makebox(-5.6,0){#1}}\end{picture}}}

\let\s\smallskip
\let\e\epsilon
\begin{document}

\title[Super-Virasoro singular vectors and Jack superpolynomials]{The super-Virasoro singular vectors and \\ Jack superpolynomials relationship revisited} 
\thanks{\today}

\author{O.~Blondeau-Fournier}
\address{Department of Mathematics, King's College London, Strand, United Kingdom, WC2R~2LS.}
\email{olivier.blondeau-fournier@kcl.ac.uk}

\author{P.~Mathieu}
\address{D\'epartement de Physique, de G\'enie Physique et
d'Optique, Universit\'e Laval,  Qu\'ebec, Canada, G1V~0A6.}
\email{pmathieu@phy.ulaval.ca}

\author{D.~Ridout}
\address{School of Mathematics and Statistics,
University of Melbourne,
Parkville, Australia, 3010.}
\email{david.ridout@unimelb.edu.au}

\author{S.~Wood}
\address{Mathematical Sciences Institute, Australian National University, Acton, Australia, 2601.}
\email{simon.wood@anu.edu.au}

 \begin{abstract}
	A recent novel derivation of the representation of Virasoro singular vectors in terms of Jack polynomials is extended to the supersymmetric case.  The resulting expression of a generic super-Virasoro singular vector is given in terms of a simple differential operator (whose form is characteristic of the sector, Neveu-Schwarz or Ramond) acting on a Jack superpolynomial. The latter is indexed by a superpartition depending upon the two integers $r,s$ that specify the reducible module under consideration. The corresponding singular vector (at grade $rs/2$), when expanded as a linear combination of Jack superpolynomials, results in an expression that (in addition to being proved) turns out to be more compact than those that have been previously conjectured. 	As an aside, in relation with the differential operator alluded to above, a remarkable property of the Jack superpolynomials at $\alpha=-3$ is pointed out.
\end{abstract}

\maketitle

\section{Introduction}
\subsection{Brief overview of past results in the superconformal case}
The remarkable relationship between Virasoro singular vectors and Jack polynomials \cite{MY,AMOSa,SSAFR,RW}
turns out to have a surprising supersymmetric counterpart. Indeed, there are two completely different formulations
of this  representation of the super-Virasoro singular vectors in terms of symmetric polynomials.\\

The first one \cite{DLM_SCFT} relies on the extension of the Jack polynomials to superpolynomials, which are polynomials that depend upon  additional anticommuting variables. In this context, the simple link which exists in the Virasoro case, namely that a singular vector is represented by a single Jack indexed by a rectangular partition, becomes somewhat more complicated. Indeed, instead of a single Jack superpolynomial ({sJack} for short) indexed by a superpartition of rectangular form, a singular vector is represented by a linear combination of sJacks  indexed by the so-called self-complementary superpartitions which are controlled by the expected rectangle. A self-complementary superpartition is such that when glued in a specific manner with (a slightly modified version of) itself, it fills the defining rectangle.  The explicit expression for the coefficients of these linear combinations has been conjectured (and heavily tested) in both the Neveu-Schwarz (NS) and Ramond (R) sectors\cite{DLM_SCFT,ADM}. \\

The second one \cite{BBT}, proposed almost simultaneously, bypasses the manipulation of fermionic variables by bosonising the free fermion modes in an intricate way. Then the symmetric polynomial representing the singular vector is a Uglov polynomial indexed by a rectangular partition, exactly as in the Virasoro case.  The Uglov polynomials, like the Jacks, can be obtained from the Macdonald polynomials, which depend upon two parameters $q$ and $t$, in a special limit. While the Jack case is recovered by taking $q=t^\a$ and $t\to 1$, the {Uglov} case corresponds to $q=-|t|^\alpha$ and $ t\to -1$ (see e.g., \cite[Chap.~10]{KK}). This connection, conjectured  in \cite{BBT}, has recently been proved in \cite{Yan}.
Notice however that this second construction is limited to the NS sector.\\

In the present work, we reconsider the first approach along the lines of the argument of \cite{RW} which is sketched in the following subsection.

\subsection{Schematic version of the argument in the Virasoro case}

The derivation of the relationship Virasoro-Jack proceeds via the following steps.

\begin{enumerate}
\item Using the free boson representation of the Virasoro algebra, write the singular vector at grade $rs$ in terms of a composite screening charge as
\be
|\chi_{r,s}\R=\int dz\, {f(z,\mathfrak{h}_-)} \,|\eta_{r,s}\R
\ee
where $f$  is a polynomial in the free boson negative modes ($\mathfrak{h}_-$ is the subalgebra of the negative modes of the Heisenberg algebra, see \eqref{comaa})  and in the $z=(z_1, \ldots, z_r)$ variables.   This function comes from 
the normally ordered product of {$r$} vertex operators together with the prefactor generated by this ordering.  The vector $|\eta_{r,s}\R$ is a suitable Fock highest-weight state specified by two positive integers $r$ and $s$, such that the resulting expression $|\chi_{r,s}\R$ is a well-defined (nontrivial) singular vector.\\

\item Identify, within $f$, a Jack polynomial evaluated at $z_i^{-1}$, so that
\be f(z,\mathfrak{h}_-)=P^{(\a)}_{(s^r)}(z^{-1}) R(z,\mathfrak{h}_-)
\ee
where $(s^r)$ is the rectangular partition whose diagram has $r$ rows all of length $s$ and $\a$ is a free parameter --- free because this Jack actually does not depend upon $\a$, being equal to a single monomial.  This is so because the number of variables is equal to the number of parts in $(s^r)$ and because $(s^r)$ is the lowest partition (with respect to the dominance ordering) of degree $rs$ having this property.  The term $R(z,\mathfrak{h}_-)$ contains a factor $w(z;\kappa)$ which is precisely the weight function defining the  scalar product for the Jacks with {a particular} parameter $\kappa$:
\be \L  P^{(\kappa)}_\la,P^{(\kappa)}_\mu\R\propto \int dz \, w(z;\kappa) \, P^{(\kappa)}_\la(z^{-1})\, P_\nu^{(\kappa)}(z).\ee
This suggests trying to write the integral over ${f(z,\mathfrak{h}_-)}$ in terms of  a scalar product for which, by setting $\a=\kappa$, we already have the left component, namely $P^{(\kappa)}_{(s^r)}(z^{-1})$, and the weight {$w(z;\kappa)$}.\\

\item Introduce an algebra isomorphism ${\varrho}$ relating the operators $a_{n}$, $n<0$, of $\mathfrak{h}_-$ to symmetric polynomials: 
\be \varrho(p_m)=\gamma\, a_{-m}\ee
where $\gamma$ is  a constant to be fixed shortly  and $p_m=\sum_{i=1}^\infty y_i^m$ with the $y_i$ being new variables (in unlimited number). Under the inverse action of this isomorphism, $R(z,\mathfrak{h}_-)$ gets transformed into $R(z,y)$.\\

\item Then observe that there is a particular value of $\gamma$, related to $\kappa$, for which $R(z,y)$ can be decomposed as
\be R(z,y)={w(z;\kappa)} \sum_\nu P_\nu^{(\kappa)}(z)P_\nu^{(\kappa)}(y) \, b_\nu(\kappa)\,
 \ee
for some constant $b_\nu(\kappa)$. We have thus
\be 
|\chi_{r,s}\R=\sum_\nu     b_\nu(\kappa)\,  \varrho\l(P_\nu^{(\kappa)}(y)\r)     \int dz\, {w(z;\kappa)} \, P^{(\kappa)}_{(s^r)}(z^{-1})\, P_\nu^{(\kappa)}(z)\,|\eta_{r,s}\R
\ee 

\item We next use the orthogonality property of the Jacks, namely:
\be \int dz \, w(z;\kappa) \, P^{(\kappa)}_\la(z^{-1})\, P_\nu^{(\kappa)}(z)\,\propto \delta_{\la,\nu}\ee
to conclude that
\be 
{ |\chi_{r,s}\R}\propto  {\varrho}\bigl( P_{(s^r)}^{(\kappa)}(y) \bigr)\,|\eta_{r,s}\R
\ee
the sought-for result.
\end{enumerate}

\subsection{How the argument is modified in the superconformal case}
Again we keep the presentation at a sketchy level. The free field representation of the superconformal algebra involves a free boson and a free fermion.
\begin{enumerate}
\item  A singular vector at grade $rs/2$ (in a reducible highest-weight module of the superconformal algebra) is written in terms of a composite screening charge
\be\label{eqsupersvex1su}
|\chi_{r,s}\R=\int dz \, d\ta \, F(z,\ta, (\mathfrak h \otimes  \mathcal C)_-) |\eta_{r,s}\rangle
\ee
which is  expressed in terms of the variables $z=(z_1, \ldots, z_r)$ and a set of anticommuting  variables $\ta=(\ta_1, \ldots, \ta_r)$.  The function $F(\cdot)$ is the product of $r$ normally ordered super-vertex operators.  The notation  $(\mathfrak h \otimes  \mathcal C)_-$ refers to the negative modes  of the free boson and the free fermion ($\mathcal C$ stands for Clifford algebra, which is sector-dependent, NS or R). The highest weight   $|\eta_{r,s}\rangle$ is the tensor product of suitable bosonic and  fermionic highest-weight states.\\

\item Appealing to the theory of symmetric superpolynomials, we identify within the function $F(\cdot)$ a Jack superpolynomial $P_\Gamma^{(\kappa)}$ in the variables $(z^{-1}, \partial_{\ta})$, i.e.,
\be
F(z,\ta, (\mathfrak h \otimes  \mathcal C)_-)= P_\Gamma^{(\kappa)}(z^{-1},\partial_\ta)  \, B(z,\ta) \, w(z;\kappa)\,  \mathrm R(z,\ta, (\mathfrak h \otimes  \mathcal C)_-).
\ee
The replacement of the variables $(z,\ta)$  by $(z^{-1},\d_\ta)$ in a superpolynomial can be interpreted as the adjoint operation defined by the orthogonality relation of superpolynomials: 
   \be P_\Gamma^{(\kappa)}(z^{-1},\partial_\ta)=\l[P_\Gamma^{(\kappa)}(z,\ta)\r]^\dagger.\ee
    The label $\Ga$, called a superpartition, is a direct generalization of the rectangular partition $(s^r)$.   
   In addition, we obtain a residual (sector-dependent) factor  $B(z,\ta)$ that is neither part of a sJack nor part of the weight function $w(z;\kappa)$ defining the sJack scalar product.   The expression for 
   $B(z,\ta)$ can be transformed into an operator that acts on the Jack superpolynomial, inside the orthogonality relation:
   \be P_\Gamma^{(\kappa)}(z^{-1},\partial_\ta)  \, B(z,\ta) =\l[B^\dagger P_\Gamma^{(\kappa)}(z,\ta)\r]^\dagger .
 \ee  $B^\dagger$ is symmetric but it does not act diagonally on $P_\Gamma^{(\kappa)}(z,\ta)$. Therefore $B^\dagger P_\Gamma^{(\kappa)}(z,\ta)$ generates a linear combination of Jack superpolynomials:
 \be   B^\dagger P_\Gamma^{(\kappa)}(z,\ta)=\sum_{\Omega} d_{\Ga,\Omega}(\kappa) P_\Omega^{(\kappa)}(z,\ta).
 \ee\\

 \item
We then define an algebra isomorphism  whose inverse maps the term $\mathrm R(\cdot)$ into the power sum version of the Cauchy formula which can be transformed into a bilinear sum of sJacks:
\be
\varrho \colon   \mathrm R(z,\ta, (\mathfrak h \otimes  \mathcal C)_-)  \; \leftarrow \; \sum_\La P_\La^{(\kappa)}(z,\ta) P_\La^{(\kappa)}(y,\phi) b_\La(\kappa).
\ee

\item
The singular vector \eqref{eqsupersvex1su} becomes then 
\be|\chi_{r,s}\R=
\sum_\La  b_\La(\kappa) \, \varrho\bigl(P_\La^{(\kappa)}(y,\phi) \bigr)  \int dz \, d\ta \,  w(z;\kappa) \l[B^\dagger P_\Gamma^{(\kappa)}(z,\ta)\r]^\dagger \, P_\La^{(\kappa)}(z,\ta)\,|\eta_{r,s}\R.
\ee

\item
Finally, we use the orthogonality relation of the sJacks,
\be  \int dz \, d\ta \,  w(z;\kappa) \l[ P_\Omega^{(\kappa)}(z,\ta)\r]^\dagger \, P_\La^{(\kappa)}(z,\ta)\propto \delta_{\Omega,\La},
\ee
 to conclude that the singular vector is given  by a linear combination of the terms that appear in $B^\dagger P_\Gamma^{(\kappa)}(z,\ta)$ but whose expansion coefficients are dressed by normalization factors $n_\Omega$, i.e., $d_{\Ga,\Omega}\to d_{\Ga,\Omega}\,n_\Omega$. The resulting expression for the singular vector is  thus of the form
 \be|\chi_{r,s}\R=\sum_\Omega d_{\Ga,\Omega}\,n_\Omega\,\varrho\l( P_\Omega^{(\kappa)}(y,\phi)\r)\,|\eta_{r,s}\R\ee
Now, the factors $n_\Omega$ are known. However, in general, the coefficients $ d_{\Ga,\Omega}$ are not known.
We thus end up an implicit, albeit general, expression for the singular vectors.
\end{enumerate}

How does this compare with the  conjectured expressions in \cite{DLM_SCFT,ADM}? 
Note first that these  latter results are presented in a different sJack basis, that is, for a different value of the Jack parameter, which explains the mismatch with the present results. In general, the expressions presented here are simpler in that they contain less terms. Another advantage is that here they are derived, and therefore proved, as opposed to being conjectured. But in \cite{DLM_SCFT,ADM}, there is an explicit conjecture for the expression of all the coefficients for any singular vector, which is not the case here.

\subsection{Organization of the article}  In Section~\ref{Svirsvsc}, we collect review material on the superconformal algebra, its free field representation and the definition of {the} screening charges from which the singular vectors are constructed.
Sections \ref{NSsv} and \ref{Rsv} are devoted to the explicit derivation of the representation of the singular vectors in terms of symmetric superpolynomials, in the NS and R sectors respectively. 
These analyses rely on the theory of Jack superpolynomials, which is summarized in Appendix~\ref{APPPsJack}.  Finally, in Appendix~\ref{ED}, which is somewhat off the main theme of the article, we display a remarkable formula for the action of the operator $B$, in the NS sector, on the sJacks $P_\Ga^{(-3)}$.

\section{Super-Virasoro singular vectors from screening charges}
\label{Svirsvsc}

\subsection{The superconformal algebra}

The generators of the superconformal transformations are the stress-energy {tensor} $T(z)$ and its superpartner $G(z)$, which satisfy the following {operator product expansions} (OPEs):
\begin{align} \label{scftope1}
T(z)T(w) &\sim \frac{c/2}{(z-w)^4} + \frac{2T(w)}{(z-w)^2} + \frac{\partial T(w)}{z-w} \\
T(z) G(w) &\sim \frac{(3/2)G(w)}{(z-w)^2} + \frac{\partial G(w)}{z-w} \\
G(z) G(w) &\sim \frac{2c/3}{(z-w)^3} + \frac{2T(w)}{z-w} \label{scftope2}
\end{align}
where $c\in \mathbb C$ is 
the central charge.  Writing  the mode decomposition of these fields as
\be
T(z) = \sum_{n\in \mathbb Z} L_n z^{-n-2}, \qquad  \qquad G(z) = \sum_{n\in \mathbb Z+ \epsilon} G_n z^{-n-3/2},
\ee
where $\e$ specifies the sector
\be\label{defZ}
\epsilon=\begin{cases} \frac12\quad &\text{Neveu-Schwarz (NS)}
\\
0 \quad & \text{Ramond (R)},
\end{cases}
\ee  
the expressions \eqref{scftope1}--\eqref{scftope2} are equivalent to the (anti)commutation relations
\begin{align}
&[L_m, L_n]=(m-n) L_{m+n} + \frac1{12} c (m^3-m)\delta_{m,-n} \label{N1CFTalg1} \\
&[L_m, G_k]=( \frac12 m-k) G_{m+k}\\
&\{ G_k, G_l \} = 2 L_{k+l} + \frac13 c (k^2-\frac14) \delta_{k,-l}. \label{N1CFTalg2}
\end{align}
The relations \eqref{N1CFTalg1}--\eqref{N1CFTalg2}, with $n,m \in \mathbb Z$ and $ k,l \in \mathbb Z+\epsilon$, define  
the $N=1$ superconformal algebra which is denoted by $\mathfrak{svir}_\e$.\\

Let $M_{c,h}$  denote the Verma module freely generated from the highest-weight vector  $|h\rangle$ defined by
\be\label{eqhwvec1}
L_0 |h\rangle = h|h\rangle,  \qquad L_n |h\rangle = G_k |h\rangle =0, \quad n, k>0,
\ee
$h$ being the conformal dimension.  
A basis for the descendant vectors in $M_{c,h}$ is 
\be \label{vec1descvect}
 G_{-n_1}\cdots G_{-n_k} L_{-m_1} \cdots L_{-m_l} |h\rangle
\ee
for $k,l\geq 0$, {with} $n_1>n_2>\cdots>{n_k\geq0}$ and $m_1\geq m_2 \geq \cdots \geq m_l{> 0}$.  The grade of the vector \eqref{vec1descvect} is 
its $L_0$--eigenvalue  relative to the highest-weight state, namely
 $\sum_in_i + \sum_j m_j.$\\

For generic {values of} $c$ and $h$ the module $M_{c,h}$ is irreducible.  
But when $c$ and $h$ are related to each other in a special way the Verma module is reducible.    
More precisely,  let $c=c(t)$ and $h=h_{r,s}(t)$ be parametrized by the following expressions
\be\label{cent} c(t)= \frac{15}{2} -3\left ( t+\frac{1}{t}\right)  
\ee and
\be  \label{hrsent} h_{r,s}(t) =\frac{(r^2-1)t}{8}+\frac{(s^2-1)}{8t} +\frac{1-rs}{4}
+\frac1{32}(1-(-1)^{r+s})
\ee
 where $t \in \mathbb C$ and $r,s \in \mathbb Z$ (and $rs \in \mathbb Z_+$).  Then, the Verma module $M_{c(t), h_{r,s}(t)}$, for all $t\ne 0$, contains
 a singular vector at grade $rs/2$. We stress that the parity of $r+s$ determines the sector of $\svir_\e$: \mbox{$2\e=(r+s-1) \bmod{2}$}.

\subsection{Free field representation} \label{sec:ffr}
In view of connecting the expression for the singular vectors of the superconformal algebra with symmetric polynomials, {we first define an embedding of $\svir_\e$ into a free field algebra, namely, the algebra of} a free boson and a free fermion.  
Let $a(z)=\sum_{n\in \mathbb Z}a_n z^{-n-1}$ be the field whose Laurent modes satisfy the 
commutation relations of the Heisenberg algebra $\mathfrak{h}$,
\be\label{comaa} [a_m, a_n]= m \delta_{m,-n}. \ee
The field $a(z)$ is the derivative of the free bosonic field
\be\phi(z) =  a^*+   a_0 \log (z)  -\sum_{n\in \mathbb Z, n\neq 0} \frac {a_n}{n}z^{-n} 
\ee
It is thus natural to extend $\mathfrak{h}$ by the addition of  the mode $a^*$, which satisfies the commutation relation
\be \label{comaa*}
[a_n,a^*]=\delta_{n,0}. \ee

Verma modules for $\mathfrak{h}$ are called Fock modules.  A highest-weight state $|\la\rangle$ of the Fock module $\mathcal F(\la)$ with $\la\in \mathbb C$ is defined by
\be\label{bhws}
a_0 |\lambda \rangle = \la |\lambda\rangle, \qquad \quad a_n |\la \rangle = 0, \quad \forall n>0.
\ee
{The} Fock modules $\mathcal F(\la)$ are irreducible and, as  vector spaces, they are isomorphic to
\be
\mathcal F(\la) \cong S(\mathfrak{h}_-) = \mathbb C[a_{-1}, a_{-2}, \ldots]\,
\ee
where $S(\mathfrak{h}_-)$ is the symmetric algebra of $\mathfrak{h}_-$ or, equivalently, the polynomial ring in the negative modes of $\mathfrak{h}$.   
 Note that the exponential of the operator $a^*$ acts on  Fock modules as follows:
\be
\exp ({\mu\, a^*}) \colon \mathcal F(\la) \; \mapsto \; \mathcal F({\la+\mu}) .
\ee

Since  the superconformal algebra contains odd generators  ($G_r$, $r\in \mathbb Z + \epsilon$), we also need to introduce a Clifford algebra in  the free field representation.    
   Let $b(z) = \sum_{n\in \mathbb Z+\e}b_n z^{-n-\frac12}$ be the free fermionic field whose Laurent modes satisfy the Clifford algebra $\mathcal C_\e$,
\be\label{atcombb}
 \{b_m, b_n\} = \delta_{n,-m}
\ee
where $n,m\in \mathbb Z+\e$ and $\e = 0, \frac{1}{2}$.  There is a unique Fock (Verma) module over $\mathcal{C}_{\e}$, for each choice of $\e$, and it is likewise irreducible.  In the Neveu-Schwarz sector ($\e = \frac{1}{2}$), the fermionic Fock module is isomorphic, as a vector space, to the exterior algebra in the negative modes:
\begin{equation}
	\exterior[b_{-1/2}, b_{-3/2}, \ldots].
\end{equation}
\\

The Ramond sector presents a minor complication because the fermionic Fock module has two independent ground states that are interchanged by $b_0$.  As $b_0$ squares to $\frac{1}{2}$ and not $0$, the Ramond Fock module cannot be identified with $\exterior[b_0, b_{-1}, \ldots]$.  It may be realized as a direct sum of two copies of $\exterior[b_{-1}, b_{-2}, \ldots]$, however we shall find it convenient to instead follow \cite[App.~B]{DLM_SCFT} and decompose the zero mode as
\be
b_0 = \frac{1}{\sqrt{2}}(b_0^+ + b_0^-).
\ee
Here, $b_0^+$ annihilates the highest-weight state of the Ramond Fock module and $b_0^-$ maps it to the other independent ground state.  Imposing $(b_0^\pm)^2=0$ and $\{b_0^+, b_0^- \}=1$ ensures the validity of the identity $b_0^2=\frac12$ since
\be
b_0^2=\frac{1}{2}(b_0^+)^2 + \frac{1}{2} \{b_0^+, b_0^-\} + \frac{1}{2} (b_0^-)^2 .
\ee
Moreover, we may also impose $\{b_0^{\pm}, b_n\}=0$ for all $n \neq 0$.  With this decomposition, we realize the Ramond Fock module as being isomorphic, as a vector space, to the exterior algebra
\begin{equation}
	\exterior[b_0^-, b_{-1}, b_{-2}, \ldots].
\end{equation}
\\

The full free field algebra is the tensor product of $\mathfrak{h}$ and $\mathcal{C}_\e$. 
The corresponding Fock modules, denoted by $\mathcal F_\e(\la)$, are the tensor product of the Fock module $\mathcal{F}(\la)$ for $\mathfrak{h}$ with either the Neveu-Schwarz Fock module ($\e=\frac{1}{2}$) or the Ramond Fock module ($\e=0$).  They are characterized by their highest-weight states which satisfy \eqref{bhws} as well as
\be
b_0^+ |\la\rangle = 0, \quad \e=0; \qquad
b_n|\la\rangle = 0, \quad \forall n>0.
\ee
Descendant states in  $\mathcal F_\e(\la)$ are of the form
\be
 a_{-n_1} \cdots a_{-n_k} b_{-m_1} \cdots b_{-m_l} |\la\rangle
\ee
for $k,l\geq 0$ with $n_1\geq \cdots \geq n_k>0$, $m_1>\cdots > m_l\geq 0$,  and $n_i\in \mathbb Z, m_j\in \mathbb Z+\e$.  As a vector space, it is clear that
\begin{equation}
	\mathcal{F}_{\e}(\la) \cong
	\begin{cases*}
		\mathbb{C}[a_{-1}, a_{-2}, \ldots] \otimes \exterior[b_{-1/2}, b_{-3/2}, \ldots] & if \(\e = \frac{1}{2}\), \\
		\mathbb{C}[a_{-1}, a_{-2}, \ldots] \otimes \exterior[b_0^-, b_{-1}, b_{-2} \ldots] & if \(\e = 0\).
	\end{cases*}
\end{equation}
\\

The relations \eqref{comaa}, \eqref{comaa*} and \eqref{atcombb} imply the following OPEs
 \be
  \phi(z)  \phi(w) \sim \log(z-w), \qquad b(z)b(w) \sim \frac{1}{(z-w)}.
\ee
Note that the fermionic two-point function is sector-dependent
\be \label{OPEbbNSR}\begin{split}
\L b(z)b(w) \R=\frac{1}{(z-w)}  \qquad \text{(NS)} \qquad\text{and}\qquad
\L b(z)b(w) \R = \frac{1}{2}\frac{\sqrt{z/w}+\sqrt{w/z}}{(z-w)} \qquad \text{(R)} .\end{split}
\ee
The sought for realization of the superconformal generators is
\be\label{freefieldsTG1}\begin{split}
T(z) &= \frac12 ( \partial \phi(z) \partial \phi(z) ) + \frac{\a_0}{2} \partial^2\phi(z) + \frac12 ( \partial b(z) b(z) )\\
G(z) &= \partial\phi(z)b(z) + \a_0 \partial b(z)
\end{split}
\ee 
where $\alpha_0$ is a free (complex) parameter (the background charge in  the Coulomb gas formalism) related to the central charge by
\be
c=\frac32 - 3 \a_0^2.
\ee
In terms of modes, \eqref{freefieldsTG1} becomes
\begin{align}\label{ffLG}
L_n&=-\frac12\a_0 (n+1)a_n+\frac{1}{2}\sum_{m\in\mathbb{Z}}a_ma_{n-m}+\frac{1}{2}\sum_{k\in\mathbb{Z}+{\epsilon}}\l(k+\frac{1}{2}\r)b_{n-k}b_{k} \quad (n\ne0),\nonumber
\\
L_0&=\frac12(a_0^2-\a_0a_0)+
\sum_{\substack{m\in\mathbb{Z}\\ m>0}}a_{-m}a_{m}+\sum_{\substack{k\in\mathbb{Z}+{\epsilon}\\k>0}}
k\, b_{-k}b_{k} +\frac{(1-2\e)}{16},\nonumber
\\
G_k&=-\a_0 \left(k+\frac{1}{2}\right)b_k+\sum_{m\in\mathbb{Z}}a_mb_{k-m},
\end{align}
where, we recall, $\epsilon=0 \,(\tfrac12)$ in the R (NS) sector.  
\\

Using the above expression for $L_0$,
the conformal dimension of a highest-weight state in $\mathcal F_\e(\la)$
is found to be
\be\label{confodimla}
 L_0 |\la \rangle =  \l( h_\la +\frac{(1-2\e)}{16} \r) |\la\rangle
\ee
where
\be \label{defhla}h_\la = \frac12(\la^2- \alpha_0 \la). \ee

\s
\subsection{Vertex operators and screening charges}
In terms of the free fields just introduced, we define the (super) vertex operator as
\be\label{SuperVertexeq1}
\mathsf V_\la(\zeta) =  \; : \exp \la [ \phi(z) + \ta b(z) ] \, :  \ee
where  $\zeta=(z,\ta)$ and $\ta$ is a Grassmann variable. The normal ordering in the vertex operator means that the positive and negative free field modes are separated, the former being placed at the right.  
  By a straightforward computation, the normal ordering for two vertex operators, with $\zeta_1=(z_1, \ta_1)$ and $\zeta_2=(z_2, \ta_2)$, is found to be
\be\label{normORVNS}
\mathsf V_\la(\zeta_1) \mathsf V_\mu (\zeta_2) = (z_1-z_2)^{\la \mu } \l( 1- \la \mu \frac{\ta_1 \ta_2}{z_1-z_2} \r)  \; : \mathsf V_\la(\zeta_1) \mathsf V_\mu (\zeta_2) : 
\ee
in the NS sector and
\be\label{VV}
\mathsf V_\la(\zeta_1) \mathsf V_\mu (\zeta_2) = (z_1-z_2)^{\la \mu} \l(1- \la \mu \frac{\ta_1 \ta_2 }{2\sqrt{z_1 z_2}} \; \l( \frac{z_1+z_2}{z_1-z_2} \r) \r)  : \mathsf V_\la(\zeta_1) \mathsf V_\mu (\zeta_2) : 
\ee
in the R sector. \\

It is also simple to verify that $\mathsf V_\la(\zeta)$ is a primary superfield, whose decomposition  into
field components takes the form
\be \mathsf V_\la(\zeta) =  \; : \exp [\la  \phi(z)]: +\, \ta\, \la\, b(z) \, : \exp [\la  \phi(z)]:.  \ee
For
$\Phi(\zeta)$ to be a  primary superfield of dimension $h$, with $\Phi(\zeta) =
\varphi (z)+
\ta
\xi (z)$,
the following conditions need to be satisfied:
\begin{align}\label{superpri}
T(z) \varphi(w) &\sim \frac{h \varphi(w)}{(z-w)^2} + \frac{\d  \varphi(w)}{z-w}&T(z) \xi(w) &\sim \frac{(h+\frac12)\xi(w)}{(z-w)^2} + \frac{\d \xi(w)}{z-w}\nonumber\\
G(z)  \varphi(w) &\sim \frac{\xi(w)}{(z-w)}&
G(z) \xi(w) &\sim \frac{2h \varphi(w)}{(z-w)^2} + \frac{\d  \varphi(w)}{z-w}.
\end{align}
The conformal dimension of the field $\mathsf V_\la(\zeta)$ is  given by $h_\la$, {see} 
\eqref{defhla}. 
 This entails the correspondence
\be
\mathsf V_\la(\zeta)\,{|0\R} \; \leftrightarrow \; |\la\rangle .
\ee

We next introduce the screening charges as\be
Q_{\pm}= \int dz\,d\ta \; \mathsf V_{\a_\pm}(\zeta)
\ee
(with the convention $\int d\,\ta \,\ta^p=\delta_{p,1}$) where the value of $\alpha_\pm$ is fixed by enforcing $h_{\a_\pm}=1/2$, which yields 
\be\label{eqalphapm12}
\a_\pm = \frac{\a_0\pm\sqrt{\a_0^2+4}}{2}, \qquad \a_++\a_-=\a_0, \qquad \a_+\a_-=-1.
\ee
The constraint on the conformal dimension ensures the commutativity of $Q_\pm$ with both $T(z)$ and $G(z)$, which is manifest from the second and fourth OPEs in eq.~\eqref{superpri}.\\

The screening charges $Q_\pm$ define thus intertwiners between representations of the superconformal algebra.
As a result, $Q_\pm |\chi\rangle$ is a highest-weight state if $|\chi\rangle$ is itself a highest-weight state.
More generally, we can define a new set of screening charges by composition of  the screening charge $Q_\pm$.  Let $k$ be a positive integer and define 
\be\label{kcompoSC}
Q_\pm^{[k]} =  \int_{ \mathrm{C}_k} dz_1 \cdots dz_k \int d\ta_1 \cdots d\ta_k \mathsf V_{\alpm}(\zeta_1) \cdots  \mathsf V_{\alpm}(\zeta_k) 
\ee 
where $\mathrm{C}_k$ is a certain integration  contour.%
\footnote{The description of the contour $\mathrm C_k$ is somewhat subtle and we refer the reader to \cite{KatoMat, Kenji} for an explicit construction.  
For the present calculations, we will take for granted that, up to a certain normalization (which is irrelevant for our purpose), the integration contour $\mathrm C_k$  is equivalent to that appearing in the  scalar product of the Jack (super)polynomials --- see Appendix~\ref{APPPsJack}.}\\

In order to describe the superconformal singular vectors in terms of a Fock-space construction, we need to introduce the following Fock highest-weight states $|\a_{r,s}\R$:
\be
\a_{r,s}= \frac{(1-r)}{2} \a_+ + \frac{(1-s)}{2} \a_-,
\ee
with $r,s \in \mathbb Z$. 
We will denote by $\mathcal F_{r,s}=\mathcal F_\e(\alpha_{r,s})$ the Fock module associated with the highest-weight state $|\alpha_{r,s}\rangle$ with the understanding that
\be
\mathcal F_{r,s}= \begin{cases} \mathcal F_{\frac12}(\alpha_{r,s}) \qquad &\text{if} \; r+s=0 \; \bmod{2} \\
\mathcal F_{0}(\alpha_{r,s}) \qquad &\text{if} \; r+s=1 \; \bmod{2}.
\end{cases}
\ee
 The action of $ Q_\pm^{[k]}$ relates Fock modules as follows:
\be \begin{split}
& Q_+^{[r]} \, : \, \mathcal F_{r,-s} \; \mapsto \; \mathcal F_{-r,-s},
 \\
 &
  Q_-^{[s]} \, : \, \mathcal F_{-r,s} \; \mapsto \; \mathcal F_{-r,-s}.
\end{split}
\ee

The following statement links the composite screening charges with the explicit expressions for the singular vectors in $\svir_\e$. 
\begin{proposition}[\cite{KatoMat}]\label{propsv1}   In the Verma module $M_{c(t),h_{r,s}(t)}$, $rs \in \mathbb{Z}_+$, which belongs to the NS sector when $r+s$ is even and the R sector when $r+s$ is odd, there exists a nonvanishing singular vector at grade $rs/2$ given either by
\be
\label{xrs}
|\chi_{r,s}^+ \rangle =  Q_+^{[r]} |\alpha_{r,-s} \rangle, \quad{\text{for}}\; r\leq s, \quad \text{or} \quad
|\chi_{r,s}^- \rangle =  Q_-^{[s]} |\alpha_{-r,s} \rangle, \quad{\text{for}}\; r\geq s,
\ee
with $h_{r,s}=(\a_{r,s}^2-\a_0\a_{r,s})/2$.
\end{proposition}

\section{The Neveu-Schwarz algebra}
\label{NSsv}

{We first consider explicit expressions for the singular vectors in the NS sector, i.e. for the $\svir_{\frac12}$ algebra.}

\subsection{Manipulating the integral representation of the singular vectors:
  identifying therein a Jack superpolynomial}

Using the expression \eqref{normORVNS} for the normal ordering of $k$ vertex operators, one obtains for the screening operator \eqref{kcompoSC} 
\be\label{compVOforNSs123} \begin{split}
  Q_\pm^{[k]} &=  \int dz_1 \cdots dz_k \int d\ta_1 \cdots d\ta_k  \; {e^{k \alpm  \,  a^*}} \prod_{\mathclap{1\le i\neq j\le k}}\: (z_i-z_j - \ta_i \ta_j)^{\alpm^2/2} \prod_{i=1}^k z_i^{\alpm a_0} \\
& \quad\times \exp \Bigl( \alpm \sum_{n>0}  \Bigl[  \tfrac1n  p_n(z)\, a_{-n}  + \tilde p_{n-1}(z,\ta)\,b_{\frac12-n}  \Bigr] \Bigr)
\\& \quad\times \exp \Bigl( -\alpm \sum_{n>0}  \Bigl[  \tfrac1n  p_n(z^{-1}) \,a_n  - \tilde p_{n}(z^{-1},\ta)\,b_{n-\frac12}  \Bigr] \Bigr),
\end{split}
\ee
where $p_n(z)$ and {$\tilde p_n(z,\ta)$} are the power sum symmetric superpolynomials (defined in \eqref{pobasis}).  
(Note that, since the normalization of singular vectors is arbitrary, we will not care about global phase factors.) Acting with $Q_\pm^{[k]}$ on a generic highest-weight state $|\a_{p,q}\R$ of the Fock module $\mathcal F_{p,q}$, we have 
\be  \label{expregenhwvno1}\begin{split}
  Q_\pm^{[k]}|\a_{p,q}\R&= \int dz_1 \cdots dz_k \int d\ta_1 \cdots d\ta_k  \;
  \prod_{i=1}^k z_i^{\alpm \a_{p,q} + (k-1) \alpha_\pm^2/2} \: \prod_{\mathclap{1\le i\neq j\le k}}\:\ \l(1-\frac{z_i}{z_j}\r)^{\alpha_\pm^2/2} \\
  & \quad \times
   \prod_{\mathclap{1\le i\neq j\le k}}\:\ \l(1- \frac{ \alpha_\pm^2 \ta_i \ta_j}{2(z_i-z_j)} \r)  
  \exp \Bigl( \alpm \sum_{n>0}  \Bigl[  \tfrac1n  p_n(z)\, a_{-n}  + \tilde p_{n-1}(z,\ta)\,b_{\frac12-n}  \Bigr] \Bigr) |\a_{p,q}+k\alpha_{\pm}\R
\end{split}
\ee
where all positive modes have annihilated the highest-weight state. Notice that this action preserves the sector: $p+q=p+q+2k \bmod{2}$.\\

Using Prop.~\ref{propsv1}  and considering first the case of $|\chi_{r,s}^+\rangle$, this last expression specializes to 
\be \label{expxrs0}  \begin{split}
 |\chi_{r,s}^+ \rangle  &= \int \frac{dz_1}{z_1} \cdots \frac{dz_r}{z_r} \int d\ta_1 \cdots d\ta_r \,  \prod_{\mathclap{1\le i\neq j\le r}}\:\ \Bigl( 1-\frac{z_i}{z_j} \Bigr)^{\alpha_+^2/2} \prod_{i=1}^r z_i^{-(s-1)/2}\\
&\quad \times \prod_{\mathclap{1\le i\neq j\le r}}\:\ \Bigl(1-\frac{\alpha_+^2 \, \ta_i \ta_j}{2(z_i-z_j)} \Bigr)  \exp \Bigl( \alpha_+ \sum_{n>0}  \Bigl[  \tfrac1n  p_n(z) a_{-n}  + \tilde p_{n-1}(z,\ta)b_{\frac12-n}  \Bigl] \Bigr) |\alpha_{-r,-s} \rangle .
 \end{split}
\ee
The integration over the anticommuting variables is equivalent  (up to a global sign) to
\be \label{intdtaenintdtata} \int d\ta_1 \cdots d\ta_r \, = \, \int d\ta_1 \ta_1 \cdots d\ta_r \ta_r \; \partial_{\ta_1} \cdots \partial_{\ta_r}
\ee
since both operators pick up the term whose $\ta$-dependence is precisely $\ta_1 \cdots \ta_r$.  
Hence, this allows us to write
\be \label{expxrs1}  \begin{split}
 |\chi_{r,s} ^+\rangle  &= \int \frac{dz_1}{z_1} \cdots \frac{dz_r}{z_r} \int d\ta_1 \ta_1 \cdots d\ta_r \ta_r \,  \partial_{\ta_1} \cdots \partial_{\ta_r} \prod_{\mathclap{1\le i\neq j\le r}}\:\ \Bigl( 1-\frac{z_i}{z_j} \Bigr)^{\alpha_+^2/2} \prod_{i=1}^r z_i^{-(s-1)/2} \\
&\quad \times \prod_{\mathclap{1\le i\neq j\le r}}\:\ \Bigl(1-\frac{\alpha_+^2 \, \ta_i \ta_j}{2(z_i-z_j)} \Bigr)  \exp \Bigl( \alpha_+ \sum_{n>0}  \Bigl[  \tfrac1n  p_n(z) a_{-n}  + \tilde p_{n-1}(z,\ta)b_{\frac12-n}  \Bigr] \Bigr) |\alpha_{-r,-s} \rangle .
 \end{split}
\ee
With insight, one can manipulate the integrand of the last equation to obtain
\be \label{expxrs2}  \begin{split}
 |\chi_{r,s} ^+\rangle  &= \int \frac{dz_1}{z_1} \cdots \frac{dz_r}{z_r} \int d\ta_1 \ta_1 \cdots d\ta_r \ta_r \:\prod_{\mathclap{1\le i\neq j\le r}}\:\ \Bigl( 1-\frac{z_i}{z_j} \Bigr)^{(\alpha_+^2-1)/2}  \partial_{\ta_1} \cdots \partial_{\ta_r}  \Delta(z_1^{-1}, \ldots, z_r^{-1}) \prod_{i=1}^r z_i^{-(s-r)/2} \\
&\quad \times \prod_{\mathclap{1\le i\neq j\le r}}\:\ \Bigl(1-\frac{\alpha_+^2 \, \ta_i \ta_j}{2(z_i-z_j)} \Bigr)  \exp \Bigl( \alpha_+ \sum_{n>0}  \Bigl[  \tfrac1n  p_n(z) a_{-n}  + \tilde p_{n-1}(z,\ta)b_{\frac12-n}  \Bigl] \Bigr) |\alpha_{-r,-s} \rangle 
 \end{split}
\ee
where $\Delta(z_1, \ldots, z_r)$ is the Vandermonde determinant (defined in \eqref{van}).   Observe that 
with $r+s$ even the power of $z_i$ in the last expression is an integer, ensuring a well-defined integration.\\

For $r\leq s$, $r+s=0 \bmod{2}$, define the following superpartition with exactly $r$ parts 
\be\label{eqspart1Lars56}
\Ga_{r,s}:=\l( {\frac{s+r}{2}-1,\frac{s+r}{2}-2, \ldots, \frac{s-r}{2}} ; \, \r).\ee
Notice that 
\be\Ga_{r,s}\vdash (\tfrac12r(s-1)|r)
\ee
(recall the definition \eqref{defdeg}).
The Jack superpolynomial associated to this superpartition is simply
\begin{align}\label{sjacksimpleno1}
P_{\Ga_{r,s}}^{(\kappa)}(z_1, \ldots, z_r, \ta_1, \ldots, \ta_r)&= m_{\Ga_{r,s}}(z_1, \ldots, z_r, \ta_1, \ldots, \ta_r) \nonumber\\&= \ta_1\ldots \ta_r \,\Delta(z_1,\ldots, z_r) \prod_{i=1}^r z_i^{(s-r)/2}.
\end{align}
Indeed, the monomial expansion of this sJack contains a single term since all lower terms with respect to the dominance order contain more than $r$ parts and hence must be zero when the number of variables of each type is fixed to $r$, that is, when restricted to the non-zero variables $(z_1, \ldots, z_r, \ta_1, \ldots, \ta_r)$ (see eqs~\eqref{defiN} and \eqref{imenvarsmon2}).  Note that this truncated sJack is independent of the value of $\kappa$.  
\\

Thus, in the expression that appears in the first line of \eqref{expxrs2}, one recognizes the adjoint expression of this Jack superpolynomial  \eqref{sjacksimpleno1} (namely,  \eqref{sjacksimpleno1} with the replacements  $z_i\to z_i^{-1}$ and $\ta_i\to\d_{\ta_i}$ --- cf. the definition \eqref{adj}).  A closer look shows that the resulting integral has a structure akin to the (integral) scalar product defined in Appendix~\ref{APPPsJack}.  
   In the next subsection this will be made explicit.

\subsection{Implementing the relationship between free fields and symmetric polynomials}
The idea is now to make precise the above observation and  rewrite the expression \eqref{expxrs2}  in the form of a scalar product of symmetric superpolynomials, following the analysis of the Virasoro and $\widehat {\mathfrak {sl}}(2)$  cases \cite{RW,RWsl2}.
{Using equation \eqref{aspct1}, { we can rewrite}  equation \eqref{expxrs2} in the form of a  
$(\mathfrak{h}\otimes \mathcal C_\e)$-valued} scalar product as follows 
\be\label{chip1rssp1}
|\chi^+_{r,s} \rangle =   \bigl\langle P^{(\kappa)}_{\Ga_{r,s}} , F \bigr\rangle^{\kappa_+}_r \; |\alpha_{-r,-s} \rangle,
\ee
where $\kappa_+$ is fixed in order for  $(1-z_i/z_j)^{(\a_+^2-1)/2} $ to be the kernel of the scalar product, namely $(1-z_i/z_j)^{1/\kappa_+} $ (cf. \eqref{aspct1}):
\be \label{kappa1NS} 
\kappa_+=\frac2{\a_+^2-1}=\frac{2}{\a_+\a_0},
\ee
and $F$ stands for
\be\label{termFinspno1}
F= \:\prod_{\mathclap{1\le i\neq j\le r}}\:\ \Bigl(1-\frac{\alpha_+^2 \, \ta_i \ta_j}{2(z_i-z_j)} \Bigr)  \exp \Bigl( \alpha_+ \sum_{n>0}  \Bigl[  \tfrac1n  p_n(z) a_{-n}  + \tilde p_{n-1}(z,\ta)b_{\frac12-n}  \Bigr] \Bigr).
\ee
By using the orthogonality property of the Jack superpolynomials, if we set $\kappa= \kappa_+$, then the singular vector will be given by the coefficient of $F$ that is proportional to $P_{\Ga_{r,s}}^{(\kappa_+)}(z_1, \ldots, z_r, \ta_1, \ldots, \ta_r)$.  \\

Let us now introduce  a sector-dependent isomorphism $\rho_{\e}$, which, in the NS sector ($\e=\tfrac12$) is defined as
\begin{gather}\label{defrho}
\rho_{\frac12} \colon \mathbb{C}[p_1(y), p_2(y), \ldots ] \otimes \exterior [ \tilde{p}_0(y, \phi), \tilde{p_1}(y, \phi), \ldots]
\;   \to \; \mathbb{C}[a_{-1}, a_{-2}, \ldots ] \otimes \exterior [ b_{-\frac12}, b_{-\frac32}, \ldots] ,\nonumber\\
\rho_{\frac12}( p_n(y)) =  \frac{2}{\alpha_0} \;  a_{-n},\qquad\rho_{\frac12}(\tilde{p}_{m}(y, \phi)) =   \frac{2}{\alpha_0} \;  b_{-\frac12-m}.
\end{gather}
It thus relates  the symmetric superpolynomials and  the free field negative modes. For this purpose, we have introduced a new set of  (infinitely many) indeterminates $(y,\phi)=(y_1, y_2, \ldots, \phi_1, \phi_2, \ldots)$ where $y_i$  and $\phi_i$ are even and odd variables respectively.
\\

 The  expression \eqref{chip1rssp1} becomes
\be\label{exprrhom1chip1}
 |\chi^+_{r,s}\rangle =  \Bigl\langle E(r;\alpha_+) \, P^{(\kappa_+)}_{\Ga_{r,s}} , \exp\bigl(  \kappa_+^{-1} \sum_{n>0}  \Bigl[\tfrac1n p_n(z) \rho_{\frac12}(p_n(y) )+   \tilde{p}_{n-1}(z,\ta)  \rho_{\frac12}(\tilde{p}_{n-1}(y, \phi)  )  \Bigr] \bigr) \Bigr\rangle_r^{\kappa_+} |\alpha_{-r,-s} \rangle.
\ee
In this last expression, we have {swapped} the first product in $F$ given in \eqref{termFinspno1} to the left side of the scalar product, thereby replacing it by its adjoint (using the definition \eqref{adj}):
\be
E(r;\alpha_+) = \Biggl[  \quad\,\prod_{\mathclap{1\le i\neq j\le r}}\:\ \Bigl(1-\frac{\alpha_+^2 \, \ta_i \ta_j}{2(z_i-z_j)} \Bigr)  \Biggr]^\dagger= \:\prod_{\mathclap{1\leq i<j\leq r}}\:\ \Bigl(1+{\alpha_+^2} \frac{z_i \partial_{\ta_i} z_j \partial_{ \ta_j}}{z_i-z_j} \Bigr)
\ee
$E(r;\alpha_+) $ is thus an operator that acts on the Jack superpolynomial $ P^{(\kappa_+)}_{\Ga_{r,s}}$. \\  

The second term in the scalar product in \eqref{exprrhom1chip1} is nothing but the expression of the partition function (or Cauchy kernel) for Jack superpolynomials \eqref{PF}.  Actually, the normalization factors in \eqref{defrho}  were chosen in order to generate the prefactor $\kappa^{-1}_+$ in the exponential.  Hence, the  singular vector $|\chi^+_{r,s}\rangle$ takes the compact form
\be\label{xrscompl}
|\chi_{r,s}^+ \rangle =\sum_\Omega \bigl\langle E(r;\a_+) P_{\Ga_{r,s}}^{(\kappa_+)}, Q_\Omega^{(\kappa_+)}   \bigr\rangle_r^{\kappa_+} \;  \rho_{\frac12}\l(P_\Omega^{(\kappa_+)} (y,\phi)\r) \; |\alpha_{-r,-s} \rangle, \quad  r\leq s
\ee
where the sum is over superpartitions $\Om$ such that (recall the definition \eqref{defdeg}) 
\be
\Om \vdash ( \tfrac12(rs-m) \, | \, m), \qquad m=0,1,\ldots, r.
\ee
This expression \eqref{xrscompl} captures the exact form the NS singular
vector at grade $rs/2$ when $r\leq s$. 
Making it fully explicit requires the expression for the operator $E(r;\a_+)$ on $P_{\Ga_{r,s}}^{(\kappa_+)}$ expanded in the sJack basis $P_{\La}^{(\kappa_+)}$ (i.e., for the same value of the free parameter, here $\kappa_+$). This expression is not known. 
But strangely enough, this expansion in terms of the sJacks $P_{\La}^{(-3)}$ can be obtained in closed form  (see Conj. \ref{conjB1coeffPm3} in Appendix \ref{ED}). 
\\

Going back to \eqref{expregenhwvno1}, a similar analysis for the expression $|\chi_{r,s}^-\rangle$, using Prop.~\ref{propsv1}, yields
\be\label{xrscomplm}
|\chi_{r,s}^- \rangle  =\sum_\Omega \bigl\langle E(s;\a_-) P_{\Ga_{s,r}}^{(\kappa_-)}, Q_\Omega^{(\kappa_-)}   \bigr\rangle_s^{\kappa_-} \;  \rho_{\frac12}\l(P_\Omega^{(\kappa_-)} (y,\phi) \r)\; |\alpha_{-r,-s} \rangle,  \quad r\geq s,
\ee
 with $\kappa_-=2/(\alpha_0 \a_-)$ and where the sum is over superpartitions $\Om$ such that
 \be
 \Om \vdash ( \tfrac12(rs-m) \, | \, m), \qquad m=0,1,\ldots, s.
 \ee

  \subsection{Examples}
In this section we give some examples of singular vectors using the superpolynomial  construction.
We consider only cases with $r\leq  s$ and, to lighten the notation, we set
\be\label{NB}|\chi^+_{r,s} \rangle\equiv |\chi_{r,s} \rangle\qquad\text{and} \qquad \kappa_+\equiv \kappa=\frac{2}{\a_+\a_0}.\ee
In terms of  the parametrization \eqref{cent}, we have
\be\label{para}\alpha_0=\frac{(t-1)}{\sqrt{t}},\qquad \alpha_+=
\sqrt{t},\qquad \kappa=\frac2{t-1}.\ee

\subsubsection{The $|\chi_{1,s} \rangle$ sequence}
  For $s=1,3,5, \ldots\,$, we have
\be
|\chi_{1,s} \rangle= \sum_\Omega \bigl\langle  P_{ (\frac12(s-1) ; \;)  }^{(\kappa)}, Q_\Omega^{(\kappa)}   \bigr\rangle_1^{\kappa} \;  \rho_{\frac12}\l(P_\Omega^{(\kappa)} (y,\phi) \r)\; |\alpha_{-1,-s} \rangle
\ee
 since $E(1;\a_+)=1$.  Hence, only one superpartition contributes, i.e. $\Om=(\tfrac12(s-1); \,)$. We obtain 
 \be
|\chi_{1,s} \rangle\propto \rho_{\frac12}\l(P_{(  \frac12(s-1) ;\; )}^{(\kappa)} (y,\phi)\r) \; |\alpha_{-1,-s} \rangle.
\ee
Recall that any singular vector is defined up to a (non-zero) global constant.\\

{In particular}, consider the case $s=3$. The present construction yields
\begin{align}\label{cas13}
|\chi_{1,3} \rangle& \propto \rho_{\frac12}(P_{( 1 ;\; )}^{(\kappa)} (y,\phi) ) \; | \alpha_{-1,-3} \rangle\nonumber\\
&\propto[\kappa \rho_{\frac12}( p_{(1;)})+\rho_{\frac12}( p_{(0;1)})]\; | \alpha_{-1,-3} \rangle\nonumber\\
&\propto[\kappa \rho_{\frac12}(\tilde p_{1})+\rho_{\frac12}(\tilde p_0p_1)]\; | \alpha_{-1,-3} \rangle\nonumber\\
&\propto[b_{-\frac32}+\a_+
b_{-\frac12}a_{-1}]\; | \alpha_{-1,-3} \rangle\end{align}
using \eqref{defrho}.
This should be compared with the expression of the singular vector
\be |\chi_{1,3}'\rangle = 
[G_{-\frac32}-t G_{-\frac12}L_{-1}]\,|h_{1,3}\R \ee
whose free field embedding is obtained using the relations \eqref{ffLG}:
\be\begin{split}
|\chi_{1,3}' \rangle &= \bigl( a_{-1}b_{-\frac12} + a_0 b_{-\frac32} + \alpha_0 b_{-\frac32}  - t ( a_0 b_{-\frac12} + a_1 b_{-\frac32})(a_{-1}a_0) \bigr)   |\alpha_{-1,-3} \rangle
\\
&\,= \bigl( (\alpha_{-1,-3}+\alpha_0-t\alpha_{-1,-3})b_{-\frac32} + (1-t \alpha_{-1,-3}^2) b_{-\frac12} a_{-1} \bigr) |\alpha_{-1,-3} \rangle \\
& \, \propto [ b_{-\frac32} + {\beta} b_{-\frac12} a_{-1} ] \, |\alpha_{-1,-3} \rangle
\end{split}
\ee
with
\be
 {\beta}=\frac{1-t \alpha_{-1,-3}^2 }{ \alpha_{-1,-3}+\alpha_0-t\alpha_{-1,-3}} = \frac{1-(t-2)^2}{(1-t)(t-2)/\sqrt{t} + (t-1)/\sqrt{t}}  = \sqrt{t}
\ee
where we have used $\alpha_{-1,-3}=(\alpha_+^2-2)/\alpha_+$.  This confirms that $|\chi_{1,3}' \rangle = |\chi_{1,3}\rangle$.  


\subsubsection{The $|\chi_{2,s} \rangle$ sequence}
  For $s=2,4,6, \ldots\,$, we have
\be
|\chi_{2,s} \rangle= \sum_\Omega \bigl\langle  E(2;\alpha_+) P_{ (\frac s2, \frac s2-1 ; \;)  }^{(\kappa)}, Q_\Omega^{(\kappa)}   \bigr\rangle_2^{\kappa} \; \rho_{\frac12}\l( P_\Omega^{(\kappa)} (y,\phi)\r) \, |\alpha_{-2,-s} \rangle
\ee
where
\be P_{ (\frac s2, \frac s2-1 ; \;)  }^{(\kappa)} =m_{ (\frac s2, \frac s2-1 ; \;)  }=\ta_1 \ta_2\,( z_1^{s/2} z_2^{s/2-1} -z_1^{s/2-1} z_2^{s/2})=\ta_1 \ta_2 (z_1-z_2) (z_1 z_2)^{s/2-1} \ee
so that
\be\begin{split}
E(2;\alpha_+) P_{ (\frac s2, \frac s2-1 ; \;)  }^{(\kappa)} &= \Bigl(1+ \alpha_+^2 \frac{z_1 \partial_{\ta_1} z_2 \partial_{\ta_2}}{z_1-z_2}    \Bigr) \ta_1 \ta_2 (z_1-z_2) (z_1z_2)^{s/2-1} \\
&=  P_{ (\frac s2, \frac s2-1 ; \;)  }^{(\kappa)} - \alpha_+^2 P_{ (\; ; \frac s2, \frac s2 )  }^{(\kappa)}.
\end{split}
\ee
Recall that such expressions are easy to handle since we work with only two variables of each type: $z_1, z_2$ and $\ta_1, \ta_2$ (simply because $r=2$).    Hence, we can write
\begin{align}
|\chi_{2,s} \rangle&=\l[ \bigl\langle   P_{ (\frac s2, \frac s2-1 ; \;)  }^{(\kappa)}, Q_{(\frac s2, \frac s2-1 ; \;)}^{(\kappa)}   \bigr\rangle_2^{\kappa} \;\rho_{\frac12}\l(P_{ (\frac s2, \frac s2-1 ; \;)  }^{(\kappa)}(y, \phi)\r)\r.
\nonumber\\
\qquad&\l.\qquad  
-\a_+^2\,  \bigl\langle   P_{ (\; ; \frac s2, \frac s2)  }^{(\kappa)}, Q_{(\; ; \frac s2, \frac s2)}^{(\kappa)}   \bigr\rangle_2^{\kappa} \;
 \rho_{\frac12}\l(P_{ (\; ; \frac s2, \frac s2 )  }^{(\kappa)}(y, \phi)\r)\r] \, |\alpha_{-2,-s} \rangle.
\end{align}
Dividing this expression by ${\bigl\langle 1, 1  \bigr\rangle_N^\kappa}
$ and using the definition of the coefficient $c_\La(\kappa;N)$ (cf. eq.~\eqref{defc})
\be\label{defck}
c_\La(\kappa;N) = \frac{\bigl\langle P_\La^{(\kappa)} , Q_\La^{(\kappa)}  \bigr\rangle_N^\kappa }{\bigl\langle 1, 1  \bigr\rangle_N^\kappa},
\ee
then, up to an irrelevant global factor, we have
\begin{align}
|\chi_{2,s} \rangle&=\l[  \;\rho_{\frac12}\l(P_{ (\frac s2, \frac s2-1 ; \;)  }^{(\kappa)}(y, \phi)\r)
-\a_+^2\,\frac{c_{  (\; ; \frac s2, \frac s2 )  }(\kappa;2) }{c_{(\frac s2, \frac s2-1 ; \;) }(\kappa;2)  } \, \rho_{\frac12}\l(P_{ (\; ; \frac s2, \frac s2 )  }^{(\kappa)}(y, \phi)\r)\r] \, |\alpha_{-2,-s} \rangle.
\end{align}
It only remains to evaluate a ratio of coefficients $c_\La$.
Using formula \eqref{normb} and the trick mentioned at the end of Appendix~\ref{APPPsJack} (see \eqref{aspct1a} and the example given there), we obtain
\be
c_{  (\; ; \frac s2, \frac s2 )  }(\kappa;2)= \frac{1}{\kappa s/2+1} \prod_{i=1}^{s/2} \frac{\kappa(s/2-i)+2}{\kappa(s/2-i+1)}
\ee
and
\be
c_{(\frac s2, \frac s2-1 ; \;) }(\kappa;2) = 2 \kappa^{-2} \frac{\kappa+2}{\kappa s/2+1}
\prod_{i=1}^{s/2-1} \frac{\kappa(s/2-i+1)+2}{\kappa(s/2-i)}.
\ee
After simplification, we finally have
\be
|\chi_{2,s} \rangle = \l[ \rho_{\frac12}\l(P_{ (\frac s2, \frac s2-1 ; \;)  }^{(\kappa)}(y, \phi)\r) - \frac{4\alpha_+^2 \kappa  }{  s(\kappa s+4)  } \rho_{\frac12}\l(
P_{ (\; ; \frac s2, \frac s2 )  }^{(\kappa)}(y, \phi)\r)\r]\, |\alpha_{-2,-s} \rangle.
\ee
As a consistency check, we have verified explicitly the correctness of the resulting expression of the singular vector 
for the first few values of $s$.


\section{The Ramond algebra} 
\label{Rsv}

We now turn to the construction of singular vectors in the Ramond sector, i.e.~for the $\svir_{0}$ algebra. This is somewhat more complicated than for the NS case essentially because  the fermionic field $b(z)$ now decomposes into integer modes: this brings  half-integer powers of $z$ and zero modes.   
The first point spoils the single-valued character of  the vertex operator defined in \eqref{SuperVertexeq1} (which is also reflected in the normal ordered relation \eqref{VV}). We have already, in Section \ref{sec:ffr}, dealt with the second point by decomposing the zero mode $b_0$ into a linear combination of two new modes $b_0^+$ and $b_0^-$ which we treat as positive and negative, respectively.  In any case, the screening charge construction can still be applied to obtain a symmetric polynomial representation of the singular vectors.   

\subsection{Treating the square roots}
The first step in the construction of the singular vectors consists in evaluating the composite screening charge $ Q_{\pm}^{[k]}$.  We first point out a simple trick to remove the half-integer powers of the $z_i$ variables. 
Since the variables $\ta_i$ are {being} integrated, we can make the change of  variables: $\ta_i= \sqrt{z_i} \eta_i$ with $\eta_i$ being new Grassmann variables.   In this case, the Berezin integration (which, we recall, acts as a derivative) becomes
\be\label{jaco}
\int d \ta_1 \cdots d\ta_k = \int d\eta_1 \cdots d\eta_k \prod_{i=1}^k z_i^{-1/2}.
\ee
This transformation removes all the square roots of the $z_i$ in the vertex operators themselves,
since
\be\sum_{i=1}^k\ta_ib(z_i)=\sum_{i=1}^k\sum_{n\in\mathbb Z}\ta_i\, b_n\, z_i^{-n-\frac12} = \sum_{i=1}^k\sum_{n\in\mathbb Z}\eta_i \,b_n\, z_i^{-n},\ee
 and in the
prefactor resulting from their normal ordering (cf. eq.~\eqref{VV}), i.e.
\be \prod_{\mathclap{1\leq i<j\leq k}}\:\ \l(1-\frac{\a_\pm^2}{2}\frac{\ta_i\ta_j}{\sqrt{z_iz_j}}\l(\frac{z_i+z_j}{z_i-z_j}\r)\r) = \:\prod_{\mathclap{1\leq i<j\leq k}}\:\ \l(1-\frac{\a_\pm^2}{2}\,\eta_i\eta_j\,\l(\frac{z_i+z_j}{z_i-z_j}\r)\r)
\ee
leaving only the overall multiplying factor resulting from \eqref{jaco} that will enter in the construction of  the symmetric superpolynomial. We then obtain, after relabeling $\eta\rightarrow \ta$, the following expression:
\begin{align}\label{eqqpmkcompSR1}
 Q_{\pm}^{[k]} = &\int dz_1 \cdots dz_k \int d\ta_1 \cdots d\ta_k \prod_{i=1}^k z_i^{-\frac12} e^{k\alpha_{\pm}  a^*} \prod_{\mathclap{1\le i\neq j\le k}}\:(z_i-z_j)^{\alpha_{\pm}^2/2} 
 \:\prod_{\mathclap{1\leq i<j\leq k}}\:\ \Bigl( 1- \frac{ \alpha_{\pm}^2 \ta_i \ta_j}{2 }  \bigl( \frac{z_i+z_j}{z_i-z_j} \bigr)  \Bigr) \prod_{i=1}^k z_i^{\alpha_{\pm}a_0} \nonumber\\
&\qquad \times \exp\Bigl( \alpha_{\pm} \Bigl[ \sum_{n>0}\tfrac1{n} p_n(z) a_{-n}+ \tfrac{1}{\sqrt{2}} \tilde p_0(z,\ta) b_0^- + \sum_{n>0} \tilde p_n(z,\ta) b_{-n} \Bigr] \Bigr) \nonumber\\
& \qquad \times \exp\Bigl( -\alpha_{\pm} \Bigl[ \sum_{n>0}\tfrac1{n} p_n(z^{-1}) a_{n}- \tfrac{1}{\sqrt{2}} \tilde p_0(z,\ta) b_0^+ - \sum_{n>0} \tilde p_n(z^{-1},\ta) b_{n} \Bigr] \Bigr).
\end{align}

\subsection{Constructing the singular vectors}
 Using Prop.~\ref{propsv1} and eq.~\eqref{eqqpmkcompSR1},  the expression for the singular vector {$|\chi_{r,s}^+\rangle$} at level $rs/2$ is
\be\label{vsrsRS}
\begin{split}
&{|\chi_{r,s}^+\rangle} =  Q_+^{[r]} |\alpha_{r,-s} \rangle \\
&= \int \frac{dz_1}{z_1} \cdots \frac{dz_r}{z_r} \int  d\ta_1 \ta_1 \cdots d\ta_r  \ta_r \l[ \partial_{\ta_1} \cdots \partial_{\ta_r}  \, \Delta(z_1^{-1}, \ldots, z_r^{-1})
\, 
\prod_{i=1}^r z_i^{  -(s-r+1)/2 } \r] \quad\prod_{\mathclap{1\le i\neq j\le r}}\:\ \Bigl( 1- \frac{z_i}{z_j} \Bigr)^{(\alpha_+^2-1)/2} 
 \\
&\quad \times   \prod_{\mathclap{1\leq i<j\leq r}}\:\  \Bigl( 1- \frac{\alpha_+^2 \ta_i \ta_j}{2} \bigl( \frac{z_i+z_j}{z_i-z_j} \bigr)  \Bigr)   \exp \Bigl( \alpha_+ \Bigl[ \sum_{n>0}\tfrac1{n} p_n(z) a_{-n}+ \tfrac{1}{\sqrt2} \tilde p_0(z,\ta) b_0^- + \sum_{n>0} \tilde p_n(z,\ta) b_{-n} \Bigr] \Bigr) \; |\alpha_{-r,-s}\rangle.
\end{split}
\ee

In this equation, we have taken into account that the positive modes annihilate the highest-weight state.
Note that this integral is well-defined (single-valued) only if $r+s$ is odd (thanks
 to the factor $\prod_iz_i^{-1/2}$ coming from \eqref{jaco}), a condition that is satisfied in the R sector. 
 Using the  notation \eqref{eqspart1Lars56}, we introduce the superpartition  $\Ga_{r,s+1}$   of degree $(\tfrac12 rs | r)$ given by 
\be
\Ga_{r,s+1} = \l ( \frac{s+r+1}{2}-1, \frac{s+r+1}{2}-2, \ldots , \frac{s-r+1}{2} ; \, \r),
\ee
for $r<s$. The corresponding sJack is
\be
P_{\Ga_{r,s+1}}^{(\kappa)}(z_1, \ldots, z_r, \ta_1, \ldots, \ta_r) = {\ta_1} \cdots {\ta_r} 
\Delta(z_1, \ldots, z_r) \prod_{i=1}^r z_i^{(s-r+1)/2},
\ee
whose adjoint version is recognized in the first square bracket of \eqref{vsrsRS}. Again, it is independent of $\kappa$. The strategy is the same as in the NS sector. We observe that the factor
\be  \prod_{\mathclap{1\le i\neq j\le r}}\:\ \Bigl( 1- \frac{z_i}{z_j} \Bigr)^{(\alpha_+^2-1)/2} =  \:\prod_{\mathclap{1\le i\neq j\le r}}\:\ \Bigl( 1- \frac{z_i}{z_j} \Bigr)^{1/\kappa_+} ,\ee
where we used $\kappa_+=2/(\alpha_+^2-1) = 2/(\alpha_+ \alpha_0)$, can be interpreted as the kernel of a scalar product $\L \cdot \;, \; \cdot \R^{\kappa_+}_r$. To achieve this goal, we introduce the isomorphism $\rho_\e$ which, 
for $\e=0$, is defined by
\begin{gather}
\rho_0\colon {\mathbb C[p_1(y), p_2(y),  \ldots ] \otimes \exterior [ \tilde p_0(y,\phi), \tilde p_1(y,\phi) \ldots]  \; \to \; \mathbb C[ a_{-1}, a_{-2} \ldots ] \otimes \exterior [ b_0^-, b_{-1}, \ldots ],}\nonumber\\
\rho_0( p_m(y)) =  \frac{2}{\alpha_0} \;  a_{-m}, \qquad 
\rho_0(\tilde p_m(y,\phi)) = \frac{2}{\alpha_0} b_{-m},  \qquad \rho_0(\tilde p_0(y,\phi) ) = \frac{\sqrt{2}}{\alpha_0} b_0^-
\end{gather}
 for $m>0$.
 As before  $y,\phi$ stands for a new set of  variables. The exponential factor becomes then
\be \exp\biggl(  \kappa_+^{-1} \sum_{n>0}  \Bigl[\tfrac1n p_n(z) \rho_0(p_n(y) )+   \tilde{p}_{n-1}(z,\ta) \rho_0(\tilde{p}_{n-1}(y, \phi)  )  \Bigr] \biggr)\ee
and we can use the Cauchy formula \eqref{PF} to obtain
\be\label{svramzm1no2}
 |\chi^+_{r,s} \rangle = \sum_\Omega
  \bigl\langle \, D(r;\a_+) \, P_{\Ga_{r,s+1}}^{(\kappa_+)} , Q_\Om^{(\kappa_+)}  \bigr\rangle_r^{\kappa_+} \; \rho_0\l(P_\Om^{(\kappa_+)}(y,\phi) \r)\; |\alpha_{-r,-s}\rangle, \qquad r<s.
\ee
The sum is over all superpartitions $\Omega$ such that
\be
\Om \vdash ( \tfrac12rs | m  ), \qquad m=0, 1, \ldots,  r ,
\ee
and the operator $D(r;\alpha_+)$ is given by
\be
D(r;\a_+) = \l[    \quad\prod_{\mathclap{1\leq i<j\leq r}}\:\  \Bigl( 1- \frac{\alpha_+^2 \ta_i \ta_j}{2} \bigl( \frac{z_i+z_j}{z_i-z_j} \bigr)  \Bigr)   \r]^\dagger = \:\prod_{\mathclap{1\le i\neq j\le r}}\:\  \Bigl( 1+ \frac{ \alpha_{+}^2}{2}   \,  \bigl( \frac{z_i+z_j}{z_i-z_j} \bigr)   \partial_{\ta_i} \partial_{\ta_j}  \Bigr).
\ee\s

 A similar analysis for  $|\chi_{r,s}^-\rangle$ leads to the following expression
\be\label{xrspRno2}
 |\chi_{r,s}^-\rangle
=
\sum_\Om  \bigl\langle D(s;\a_-) P_{\Ga_{s,r+1}}^{(\kappa_-)} , Q_\Om^{(\kappa_-)}  \bigr\rangle_s^{\kappa_-} \;\rho_0  \l(P_\Om^{(\kappa_-)}(y,\phi) \r) \; |\alpha_{-r,-s}\rangle ,  \qquad r>s,
\ee
with $\kappa_-=2/\a_0\a_-$ and where the sum is now over superpartitions $\Omega$ such that
\be
\Om \vdash ( \tfrac12rs | m  ), \qquad m=0, 1, \ldots,  s.
\ee

\subsection{Examples}
We end the section on the Ramond algebra by presenting some examples with $r< s$, and return to the notation  \eqref{NB}.

\subsubsection{The $|\chi_{1,s} \rangle$ sequence} 
  For $r=1$ and $s=2,4,6, \ldots\,$, the operator $D$ is the identity, i.e. $D(1;\alpha_+)=1$, and thus the expression for the singular vector \eqref{svramzm1no2} contains only one term:
\be
|\chi_{1,s} \rangle = \rho_0\l(P_{(\frac s 2 ; \;)}^{(\kappa)}(y,\phi)\r)\; |\alpha_{-1,-s}\rangle.
\ee

\subsubsection{The $|\chi_{2,s} \rangle$ sequence}
For $r=2$ and $s=3,5,7,\ldots\,$, we first compute
\be\begin{split}
D(2;\alpha_+)P_{(\tfrac{s+1}{2}  , \tfrac{s-1}{2}  ; \; ) }^{(\kappa)}&= \Bigl(1+ \frac{\alpha_+^2}{2} \bigl(\frac{z_1+z_2}{z_1-z_2} \bigr) \partial_{\ta_1} \partial_{\ta_2} \Bigr)
\,\l[\ta_1 \ta_2 (z_1-z_2)z_1^{(s-1)/2}  z_2^{(s-1)/2} \r] \\
&= 
P^{(\kappa)}_{  (\tfrac{s+1}{2}  , \tfrac{s-1}{2}  ; \; ) }   - \frac{\alpha_+^2}{2}   
  P^{(\kappa)}_{  ( \; ;  \tfrac{s+1}{2}  , \tfrac{s-1}{2}  ) }.
 \end{split} 
\ee
Using eqs~\eqref{c1} and \eqref{c2}, we have
\be
c_{ ( \; ;  \tfrac{s+1}{2}  , \tfrac{s-1}{2}  )   }(\kappa;2)=\frac{2}{(1+\tfrac12\kappa(s-1)) (1+\tfrac12\kappa(s+1))} \prod_{i=1}^{\tfrac12(s-1)} \frac{2+\kappa i}{\kappa i}
\ee
and 
\be
c_{ (\tfrac{s+1}{2}  , \tfrac{s-1}{2}  ; \; )  } (\kappa;2) = 2 \kappa^{-2} 
\frac{\kappa+2}{1+\tfrac12\kappa(s+1)}
\prod_{i=1}^{\tfrac12(s-1)} \frac{\kappa(i+1)+2}{\kappa i}.
\ee
 The singular vector can be written explicitly as
\be
 |\chi_{2,s} \rangle = \l[\rho_0\l(P^{(\kappa)}_{  (\tfrac{s+1}{2}  , \tfrac{s-1}{2}  ; \; ) } (y,\phi)\r)   -\frac{\alpha_+^2}{2}   \frac{   c_{ ( \; ;  \tfrac{s+1}{2}  , \tfrac{s-1}{2}  )   }(\kappa;2)        }{     c_{ (\tfrac{s+1}{2}  , \tfrac{s-1}{2}  ; \; )  } (\kappa;2)   }\rho_0\l(  
  P^{(\kappa)}_{  ( \; ;  \tfrac{s+1}{2}  , \tfrac{s-1}{2}  ) }(y,\phi)\r)\r]\; |\alpha_{-2,-s}\rangle
\ee
and the relative coefficient can be simplified to
\be \frac{\alpha_+^2}{2} \frac{ c_{ ( \; ;  \tfrac{s+1}{2}  , \tfrac{s-1}{2}  )   }(\kappa;2)        }{     c_{ (\tfrac{s+1}{2}  , \tfrac{s-1}{2}  ; \; )  } (\kappa;2)   }= \frac{2 \alpha_+^2  \kappa^2}{ (2+\kappa(s-1)) (4+\kappa(s+1))} .\ee
We stress that this expression for $ |\chi_{2,s} \rangle$ is much simpler than that conjectured in \cite[eq.~(B.23)]{DLM_SCFT} which contains $(s+1)$ terms, instead of only  two here.

\subsubsection{A detailed evaluation of $|\chi_{2,3}\rangle$} 

Let us detail the case $(r,s)=(2,3)$ in order to further illustrate the manipulations in the R sector. 
The two contributing Jack superpolynomials are 
\be\begin{split}
P_{(2,1;)}^{(\kappa)} =& \frac{-1}{2(\kappa+1)} p_{(1,0;2)} + \frac{1}{2(\kappa+1)^2} p_{(1,0;1,1)} +\frac{\kappa}{(\kappa+1)^2} p_{(2,0;1)}  \\
& + \frac{\kappa (2\kappa+1)}{2(\kappa+1)^2} p_{(2,1;)}-\frac{\kappa}{2(\kappa+1)^2} p_{(3,0;)}
\end{split}
\ee
and
\be
P_{(;2,1)}^{(\kappa)} = \frac{-\kappa}{\kappa+2} p_{(3)} +\frac{(\kappa-1)}{(\kappa+2)} p_{(2,1)} + \frac{1}{\kappa+2} p_{(1,1,1)}.
\ee
Writing \be |\chi_{2,3}\rangle={\Xi_{2,3}}|\alpha_{-2,-3}\rangle,\ee
we have
thus (with the normalization fixed so that the coefficient of $a_{-1}^3$ is 1)
\be
\begin{split}
\Xi_{2,3} &= a_{-1}^3 -\frac{\a_0^2 \kappa}{4} a_{-3} + \frac{\a_0 (\kappa-1)}{2} a_{-2} a_{-1}
+\frac{2 (\kappa+1)(\kappa+2)}{\a_+^2 \kappa^2} b_{-1} b_0 a_{-2} - \frac{4 (\kappa+2)}{\alpha_+^2 \a_0 \kappa^2} b_{-1}b_0a_{-1}^2 \\
&\quad- \frac{4 (\kappa+2)}{\a_+^2 \kappa} b_{-2} b_0 a_{-1}  - \frac{\a_0 (2\kappa+1)(\kappa+2)}{\a_+^2 \kappa} b_{-2}b_{-1} + \frac{\alpha_0 (\kappa+2)}{\a_+^2 \kappa} b_{-3}b_0.
\end{split}
\ee
Of course, in the above expression, $b_0$ could be replaced by $ b_0^-/\sqrt{2}$.
Using $\kappa=2/(\a_0\a_+)$, $\a_0=(t-1)/\sqrt{t}$ {and} $\a_+=\sqrt{t}$, we recover 
the correct expression --- in terms of  free field modes --- for the operator $\Xi_{2,3}$ defining the singular vector $|\chi_{2,3}\R$:
\be 
\begin{split}\label{target}
\Xi_{2,3} &= a_{-1}^3  - \frac{t-1}{2{\a}} a_{-3} -  \frac{t-3}{2\sqrt{t}} a_{-2} a_{-1} + (t+1) b_{-1} b_0 a_{-2} - 2\sqrt{t} b_{-1}b_0a_{-1}^2 \\
&\quad - \frac{t+3}{\sqrt{t}} b_{-2}b_{-1} + \frac{t-1}{\sqrt{t}}b_{-3}b_0 -4 b_{-2} b_0 a_{-1}.
\end{split}
\ee

%

\section*{Acknowledgements}
OBF research is supported by a posdoctoral fellowship  from the Fonds de Recherche du Qu\'ebec, Nature et Technologie.   PM research is supported by the Natural Sciences and Engineering Research Council of Canada.  DR's research is supported by the Australian Research Council Discovery Projects DP1093910 and DP160101520.  SW research is supported by the Australian Research Council Discovery Early Career Researcher Award DE140101825 and the Discovery Project DP160101520.

\begin{appendix}

\section{Jack superpolynomials} \label{APPPsJack}

We collect in this appendix the  properties of symmetric superpolynomials used in this article.  Consider the ring
$
\mathcal R_N= {\mathbb C[z_1, \ldots, z_N] \otimes \exterior[\ta_1, \ldots, \ta_N]} 
$
in the indeterminates $z_i, \ta_i$ with $i=1, \ldots, N$.  
The elements of $\mathcal R_N$ are called superpolynomials.   
 We are interested in those elements of $\mathcal R_N$ that are fixed under the action of the symmetric group {$\mathrm S_N$} defined by $(z_i, \ta_i )\mapsto( z_{\sigma(i)}, \ta_{\sigma(i)})$, for $i=1, \ldots, N$, where $\sigma\in \mathrm S_N$. These are the  symmetric superpolynomials \cite{DLM_class}; they form a subring of $\mathcal R_N$ to be
 denoted by 
 $ \Sigma_N$.  \\

The basis elements of $\Sigma_N$ are indexed by superpartitions. 
A superpartition $\La$  is
a pair of partitions
 \begin{equation}\label{sppa}
\La=(\La^{\mathrm a};\La^{\mathrm s})=(\La_1,\ldots,\La_m;\La_{m+1},\ldots,\La_\ell),
\end{equation}
such that
\begin{equation}\label{sppb}
\La_1>\cdots>\La_m\geq0 \quad  \text{ and}
\quad \La_{m+1}\geq \La_{m+2} \geq \cdots \geq
\La_\ell > 0 \, .\end{equation}
The number of parts of $\La^{\mathrm a}$, in this case $m$, is called the fermionic degree. The bosonic degree is the sum of all the parts. A superpartition $\La$ is said to be of degree $(n|m)$, written $\La\vdash(n|m)$, if its bosonic (fermionic) degree is $n$ (resp. $m$). For easy reference, we restate this as an equation:
 \be \La\vdash(n|m)\quad \Leftrightarrow\quad n=\sum_{i=1}^\ell\La_i=|\La^a|+|\La^s|\quad\text{and}\quad m=\text{ \# of parts in}\; \La^a.
 \label{defdeg}\ee

The diagrammatic representation of $\La$ is given by the diagram of the weakly decreasing ordered partition obtained from $\La$ {by} removing the semi-colon, which is denoted by $\La^*$, and adding circles to the rows corresponding to the parts of $\La^{\mathrm a}$.  By convention, the circle is placed at the top-most row when a part of $\La^{\mathrm a}$ is repeated in $\La^{\mathrm s}$.  
Here is an example for a superpartition which has degree $(19|4)$:
\be \La=(6,3,2,0;5,3):\qquad
\tableau[scY]{&&&&&&\bl\tcercle{}\\&&&&\\&& & \bl\tcercle{}\\&&\\&&\bl\tcercle{}\\\bl\tcercle{}}
\ee
The partition $\La^*$ corresponds thus to the diagram of $
\La$ with the circles removed. The partition associated to the diagram where circles are replaced by boxes is denoted by $\La^\cd$.   
 For the above example, 
  \be \La^*=(6,5,3,3,2):\qquad  
 \tableau[scY]{&&&&&\\&&&&\\& & \\&&\\&\\}
\qquad \qquad \La^\circledast=(7,5,4,3,3,1):\qquad
\tableau[scY]{&&&&&&\\&&&&\\&& & \\&&\\&&\\ \\}
\ee
Note that the pair of partitions $\La^*$ and $ \La^\circledast$ uniquely characterizes $\La$. 
  \\

A first basis for {$\Sigma_N$} is the generalization of the monomial symmetric polynomials. Explicitly, the supermonomial 
associated to a superpartition $\La$ of fermionic degree $m$  is given by 
\begin{equation}
m_\La(z,\theta)=\theta_{1}
\cdots\theta_{m} z_1^{\Lambda_1} \cdots z_\ell^{\Lambda_\ell}+\text{distinct permutations of $(z_i,\ta_i)\lrw (z_j,\ta_j)$}.
\end{equation}
For instance, for $N=4$ we have 
\begin{align}
m_{(3,0;1,1)}(z,\theta) &=\ta_1\ta_2(z_{1}^3-z_2^3)z_3z_4+\ta_1\ta_3(z_{1}^3-z_3^3)z_2z_4+\ta_1\ta_4(z_{1}^3-z_4^3)z_2z_3\nonumber\\&\quad+\ta_2\ta_3(z_{2}^3-z_3^3)z_1z_4+\ta_2\ta_4(z_{2}^3-z_4^3)z_1z_3+\ta_3\ta_4(z_{3}^3-z_4^3)z_1z_2.
\end{align}
Observe that the parts of $\La^{\mathrm a}$ are associated to the exponents of the $z_i$ that form an antisymmetric polynomial (in the $z_i$) and the parts of $\La^{\mathrm s}$ to a symmetric polynomial.\\

The classical basis that is the most useful in the present context is that constructed multiplicatively out of the power sums: 
 \be\label{pobasis}
p_\La(z,\ta)=\tilde{p}_{\La_1}(z,\ta) \cdots \tilde{p}_{\La_m}(z,\ta) p_{\La_{m+1}}(z) \cdots p_{\La_\ell}(z).
\ee
where
 \begin{equation}\label{Spower}
 \tilde{p}_r(z,\ta)=    \sum_{i=1}^N \theta_iz_i^r\qquad\text{and}\qquad  p_s(z)=
\sum_{i=1}^Nz_i^s \, ,
\end{equation} 
with $r\geq 0$ and $s\geq 1$. 
\\

The projective limit
\be
\Sigma = \varprojlim \Sigma_N
\ee
corresponds to the case where we have infinitely many indeterminates $z_1, \ta_1, z_2, \ta_2, \ldots$.  
The monomials and power sums are stable with respect to the projective limit (meaning that the transition matrices between these two bases are independent of the number of variables, provided that the latter is large enough).
The limiting form will be denoted simply by $m_\La$ and $ p_\La$, respectively,
and are referred to as {symmetric} superfunctions.      Define $\mathfrak{i}_N$ to be the homomorphism from $\Sigma$ to $\Sigma_N$ that sets $z_{N+1}=z_{N+2}=\cdots = 0$ and $\ta_{N+1}= \ta_{N+2}=\cdots = 0$.  Then, we have 
\be\label{defiN}
\mathfrak{i}_N \bigl( m_\La \bigr) = \begin{cases}  m_\La(z_1, \ldots, z_N, \ta_1, \ldots, \ta_N)\qquad \text{if } N\geq \ell \\ 0 \quad \qquad \qquad \qquad \qquad \qquad \qquad \text{otherwise} 
\end{cases}
\ee
(recall that $\ell$ is the total number of parts in $\La$, see \eqref{sppa}).
\\

The space of symmetric superfunctions has a natural bigradation 
\be
\Sigma= \bigoplus_{n,m \geq 0}\Sigma^{(n|m)}
\ee
  where $\Sigma^{(n|m)}$ is the subspace of homogeneous symmetric superfunctions of degree $n$ in the $z_i$ and degree $m$ in the $\ta_i$. 
  It is shown in \cite{DLM_class} that
\be
\Sigma^{(n|m)} = \mathbb C m_\La, \qquad \Sigma^{(n|m)} = \mathbb C p_\La, \qquad  \quad \La \vdash (n|m).
\ee
In other words, the monomial and power sum superfunctions are independent bases over $\Sigma$.  Moreover, 
the ring $\Sigma$ is generated by the $\tilde{p}_r$ and $p_s$ with $r \ge 0$ and $s \ge 1$: 
\be
\Sigma = \mathbb C[p_1, p_2, \ldots] \otimes \exterior [\tilde p_0, \tilde p_1, \ldots].
\ee
Note that  
\be \mathfrak{i}_N:\Sigma\to \Sigma_N=\bigoplus_{n,m \geq 0}\Sigma_N^{(n|m)}.\ee

We are now in position to introduce the Jack superpolynomials.   We take the ring of symmetric superpolynomials $\Sigma_N$ defined over the field $\mathbb Q(\alpha)$ of rational functions in $\alpha$, where $\alpha$ is a free parameter.   For two elements $f,g\in \Sigma_N$ we define the following scalar product
\be\label{aspct1}
\begin{split}
& \bigl\langle \cdot, \cdot  \bigr\rangle_N^\alpha  \colon \Sigma_N \times \Sigma_N \mapsto \mathbb Q(\alpha) \\
&
\bigl\langle f,g\bigr\rangle^{\alpha}_N:= \frac{1}{(2\pi i)^N}\int_{\mathrm T} \frac{dz_1}{z_1} \cdots \frac{dz_N}{z_N} \int d\ta_1 \ta_1 \cdots d\ta_N \ta_N  \:\prod_{\mathclap{1\le i\neq j\le N}}\:\ \Bigl(1-\frac{z_i}{z_j}\Bigr)^{1/\alpha} 
 [f(z, \ta) ]^\dagger g(z,\ta)
\end{split}
\ee
where the adjoint operation with respect to this scalar product, denoted by a dagger, is given by
\be\label{adj}
[f(z,\ta)]^\dagger = f(z^{-1},\partial_\ta).
\ee
The integration in \eqref{aspct1} of the $z$ variables, with $|z_i|<1$, is over the $N$-torus $\mathrm T=\{z=(z_1, \ldots, z_N)\in \mathbb C^N\, : \, |z_i|=1,\ 1 \leq i \leq N\}$. 
\\

The Jack superpolynomials, denoted by $P_\La^{(\alpha)}(z,\ta)$, associated to superpartitions $\La$ are elements of $\Sigma_N$ defined by the  following two conditions (assuming that $N\geq \ell$). 
\begin{itemize}
\item[C1.]They decompose triangularly in the monomial basis
\be
P_\La^{(\alpha)}(z,\ta) = \sum_{\La \geq \Om} u_{\La, \Om} \,  m_\Om(z,\ta)
\ee
with $u_{\La, \Om}\in \mathbb Q(\alpha)$ {and} $u_{\La, \La}=1$. Here, $\La \geq \Om$ refers to the (partial) dominance order between superpartitions given by \cite{DLM_eval}:
\begin{equation} \label{domo}
\Lambda \geq  \Omega
\quad \text{iff}
 \quad \La^* \geq \Om^*\quad \text{and}\quad
\La^{\circledast} \geq  {\Om^{\circledast}}
\end{equation}
where the $*$- and $\cd$-type partitions are compared with the usual dominance order \cite{Mac}.
\item[C2.] They are orthogonal with respect to the scalar product \eqref{aspct1}:
\be \bigl\langle P_\La^{(\alpha)}(z,\ta),  P_\Om^{(\alpha)}(z,\ta)  \bigr\rangle_N^\alpha = 0 \qquad \text{if } \;\La \neq \Om.
\ee
\end{itemize}

The projective limit of the Jack superpolynomials, called {the} Jack superfunctions $P_\La^{(\alpha)}$, form a basis  for $\Sigma^{(n|m)}$ with all $\La\vdash (n|m)$ \cite{DLM_sJack}.
\\

Most of the properties of Jack polynomials can be lifted to the super-case  
\cite{DLM_sJack, DLM_eval, DLM_clust}.   Here we present those that are used in this work.
In order to formulate the first property in a compact notation, we introduce the dual superpolynomials $Q^{(\a)}_\La(z,\ta)$
 defined by $Q_\La^{(\alpha)}=b_\La P_\La^{(\alpha)}$ where $b_\La=b_\La(\alpha)$ is given by \cite{DLM_eval}
\be\label{normb}
b_\La=   (-1)^{\binom m2} \alpha^{-m}  \prod_{s\in \mathcal B (\La)} \frac{  \alpha \tilde{a}(s) + l(s) +1       }{ \alpha a(s) +  \tilde{l}(s)+ \alpha     }
\ee
($m$ is the fermionic degree of $\La$).  {Here}, $\mathcal B (\La)$ is the set of boxes in the diagram of $\La$ that do not lie  in a row and a column both ending with a circle.  The quantities $a(s)$ and $l(s)$ are the arm- and the leg-length calculated  in $\La^*$, while $\tilde a(s)$ {and} $\tilde l(s)$ are the analogous quantities for $\La^\cd$.
Recall that for the box $s$ in the $i$-th row and the $j$-th column of the diagram representing the partition $\la$, these data are given by
\be a(s)=\la_i-j\qquad \text{and}\qquad l(s)=\la'_j-i , \ee
where $\la'$ is the conjugate of $\la$ obtained from it by interchanging rows and columns.
We are now in position to present the three needed properties.

\begin{itemize}
\item[P1.]  \emph{The Cauchy formula (or partition function)}. We have
\be\label{PF}
 \sum_{\La} P_\La^{(\alpha)}(z,\ta) Q_\La^{(\alpha)}(y,\phi) =\exp\Bigl(  \alpha^{-1} \sum_{n>0} [\tfrac1n p_n(z) p_n(y) +   \tilde{p}_{n-1}(z,\ta)  \tilde{p}_{n-1}(y, \phi)   ] \Bigr) 
\ee
where $y, \phi$ represent another set of indeterminates
 ($y_i$ and $\phi_i$ are even and odd variables, respectively).  Note that the Cauchy formula is also valid for infinitely many variables,  i.e. in $\Sigma$.  
\\

\item[P2.] \emph{Reduction.}  For a superpartition of the form
 $\bar{\La}=(k+l-1, k+l-2, \ldots,k+1 ,k ; \,)$ which contains $l$ parts, with $k\geq 0$,  the Jack superpolynomial in exactly $l$ variables (of each type, $z$ and $\ta$) is simply  
\be \label{imenvarsmon2}
P_{\bar \La}^{(\alpha)}(z_1, \ldots, z_l,\ta_1,\ldots \ta_l) = \ta_1 \cdots \ta_l \, \Delta(z_1, \ldots, z_l) \,  m_{(k^l)}(z_1, \ldots, z_l).
\ee
Here, $\Delta(z_1, \ldots, z_l)$ denotes the Vandermonde determinant in $l$ variables,
\be\label{van}
\Delta(z_1,\ldots,z_l)=\prod_{1\leq i<j\leq l}(z_i-z_j),
\ee
and $(k^l)$ is the partition with all $l$ parts equal to $k$.  
Note in particular that this sJack does not depend upon $\alpha$.  \\

\item[P3.] \emph{Normalization}.   With respect to the scalar product \eqref{aspct1}, the normalization of a Jack superpolynomial defined by
\be\label{defc}
c_\La=c_\La(\alpha;N) := \frac{\bigl\langle P_\La^{(\alpha)} , Q_\La^{(\alpha)} \bigr\rangle_N^\alpha }{\bigl\langle 1, 1  \bigr\rangle_N^\alpha}
\ee
is conjectured to be
\be\label{c1} c_\La(\a;N)=  \alpha^{-m} {\binom N m}^{-1} c_{\La^*}(\a;N) \:\prod_{\mathclap{s\in\La^\cd\setminus \La^*}}\:(N+\a{a'(s)}-l'(s))
\ee 
with (cf. \cite[Eq.~VI~(10.37)]{Mac})
\be\label{c2} c_{\La^*}(\a;N)=\prod_{s\in\La^*}\frac{N+{a'(s)}\a-l'(s)}{N+(a'(s)+1)\a-l'(s)-1}.
\ee
In \eqref{c1} and \eqref{c2}, $a'(s)$ denotes the arm-colength of the box $s$, calculated in $\La^\cd$ and $\La^*$, respectively.  Similarly, $l'(s)$ denotes the leg-colength of $s$.  Recall that these quantities  are defined by
\be
a'(s)=j-1, \qquad l'(s)=i-1
\ee
for a box $s$ at position $(i,j)$.  
Note that the expression \eqref{defc} can be deduced from  the corresponding normalization of the Macdonald superpolynomials given in \cite[\S9]{DLM_sJack}, or \cite[\S6]{BDLM}, by taking the proper Jack limit. \end{itemize}

Finally, we end this appendix  by pointing out that instead of evaluating the integral in the scalar product \eqref{aspct1}, one can set $\a=1/k$ for $k$ a positive integer, evaluate  
\be\label{aspct1a}
\bigl\langle f,g\bigr\rangle^{1/k}_N= \ct\biggl[ \quad\,\prod_{\mathclap{1\le i\neq j\le N}}\:\ \Bigl(1-\frac{z_i}{z_j}\Bigr)^{k} (f^\dagger g)\biggr],
\ee
where $\ct[A]$ stands for the constant term of $A$, and {then} analytically continue this result to all $k\in\mathbb C$.  To illustrate this, take for example
\be
\bigl\langle1, 1 \bigr\rangle_2^{1/k}= \ct\Bigl( \bigl(1-\frac{z_1}{z_2}\bigr)^k  \bigl(1-\frac{z_2}{z_1}\bigr)^k \Bigr) =  \sum_{p=0}^k\binom{k}{p}^2=\binom{2k}{k}
\ee
and
\begin{align}
\bigl\langle P_{(1,0;)}^{(1/k)}, P_{(1,0;)}^{(1/k)} \bigr\rangle_2^{1/k}
& = (-1) \text{ct}\Bigl( \bigl(1-\frac{z_1}{z_2}\bigr)^k \bigl(1-\frac{z_2}{z_1}\bigr)^k (z_1-z_2) \bigl(\frac{1}{z_1}-\frac1{z_2}\bigr)\Bigr)
\nonumber\\& = 
(-1) \text{ct}\Bigl( \bigl(1-\frac{z_1}{z_2}\bigr)^{k+1} \bigl(1-\frac{z_2}{z_1}\bigr)^{k+1}\Bigr)= (-1)\binom{2k+2}{k+1}.
\end{align}
Hence
\be
c_{(1,0;)}(\alpha; 2)=\bigl(c_{(1,0;)}(1/k; 2)\bigr)_{k=\alpha^{-1}}     
= \alpha^{-2}  \Bigl( \frac{(2k+2)(2k+1)}{(k+1)(k+1)} \Bigr)_{k=\alpha^{-1}}
= 2\alpha^{-2} \frac{\alpha+2}{\alpha+1} .
\ee

\section{Some properties of the operators $E$ and $D$} \label{ED}

We have obtained implicit expressions for the singular vectors of the $N=1$ superconformal algebra which involve the action of the operators $E$ (for the NS sector) and $D$ (for the R sector) on Jack superpolynomials.  However, for these actions, we have not been able to find an explicit expression for their action on the Jack superpolynomial basis.    
In this section, we initiate the study of these operators and point out a surprising relation with Jack superpolynomials with the negative parameter $ \alpha=-3$, studied in \cite{DLM_clust} in  a more general setting. \\

Let $B(l;t)$ be either $E(l;\sqrt{t})$ or $D(l;\sqrt{2t})$ depending on the sector, 
\be B(l;t)=\begin{cases}  E(l;\sqrt{t})\;\,= \prod_{1\leq i<j\leq l} \Bigl(1+t  \bigl(\frac{z_i  z_j }{z_i-z_j}\bigr) \partial_{\ta_i} \partial_{\ta_j} \Bigr)\quad&\text{(NS)}\\D(l;\sqrt{2t})=\prod_{1\leq i<j\leq l}  \Bigl( 1+  t   \, \bigl( \frac{z_i+z_j}{z_i-z_j} \bigr)   \partial_{\ta_i} \partial_{\ta_j}  \Bigr)\quad\, &\text{(R).}\end{cases}
\ee 
By looking at the explicit form of the operator $B(l;t)$, we see that its action does not preserve the homogeneous space $\Sigma_l^{(n|m)}$, i.e.
\be
B(l;t) \Sigma_l^{(n|m)}  \not \subseteq \Sigma_l^{(n|m)}  .
\ee  
   It rather preserves the global even/odd parity and it decomposes as
\be
B(l;t)  \Sigma_{l}^{(n|m)} \subseteq  \Sigma_{l}^{(n|m)}
 \oplus 
 \Sigma_{l}^{(n+2\epsilon |m-2)}
\oplus 
\Sigma_{l}^{(n+4\epsilon |m-4)}
 \oplus \cdots 
\oplus
\Sigma_{l}^{(n+ 2\epsilon \lfloor \tfrac m 2 \rfloor |m \bmod{2})}
\ee
where again $\epsilon=0,1/2$ corresponds to  R, NS (respectively).  This peculiarity makes an explicit expression for the
action on sJacks difficult to obtain.   \\

For the construction of the singular vectors,  the operator $B(l;t)$ acts on a Jack superpolynomial in exactly $l$ variables which is indexed by  a staircase-shaped superpartition  
\be \label{defGab}\Gab:=\Ga_{2a-l,l}=(  a-1, a-2, \ldots, a-l ; \,   )\ee
 for a given $a\geq l$.    
Using property \eqref{imenvarsmon2} with $k=a-l$, we have
\be
P_{\Gab}^{(\alpha)}(z_1, \ldots, z_l, \ta_1, \ldots, \ta_l) = 
 m_{\Gab}(z_1, \ldots, z_l,\ta_1, \ldots, \ta_l)=\ta_1 \cdots \ta_l\, \Delta(z_1, \ldots, z_l) \, \prod_{i=1}^l z_i^{a-l}   .
\ee
 Since the operator $B(l;t)$ acts non-trivially  only  on the anticommuting variables, through their derivatives, it is convenient to expand its action on a specific monomial in the $\ta_i$ as  
\be \label{expandBenmt}
B(l;t) \ta_1 \cdots \ta_l = \sum_{k=0}^{\lfloor l/2 \rfloor} (-t)^k B_k
\ee
where we choose to write the expansion in powers of $(-t)$.  The term $B_k$ results from the action of $2k$ distinct derivatives on $\ta_1 \cdots \ta_l$, thus reducing the total fermionic degree by $2k$.  
 Because $B(l;t)$ is symmetric (in superspace), in the expansion \eqref{expandBenmt} we can  restrict ourselves to the analysis of the coefficient that contains  $\ta_1 \cdots \ta_{l-2k}$ --- the complete expression for $B_k$ then being recovered by an appropriate symmetrization.    \\

The coefficients $B_k$ turn  out to have remarkably simple compact expressions. These are\be 
{B_k}\bigl|_{\ta_1 \ldots \ta_{l-2k}} = \mathrm{pf} \Bigl( \frac{z_i z_j}{z_i-z_j} \Bigr)_{l-2k+1 \leq i \neq j \leq l}
\ee
for the NS sector and
\be
{B_k}\bigl|_{\ta_1 \ldots \ta_{l-2k}} =  \mathrm{pf} \Bigl( \frac{z_i+ z_j}{z_i-z_j} \Bigr)_{l-2k+1 \leq i \neq j \leq l}
\ee
for the R sector, where $\mathrm{pf}(\cdot)$ denotes the pfaffian.   
Using the identities \cite{BMRW2}\footnote{To make a direct comparison with \cite{BMRW2}, we write
\[ \mathrm{pf}  \left( \frac{z_i z_j}{z_i-z_j} \right)_{1\leq i \neq j \leq 2n}=\prod_{i=1}^{2n}z_i\: \mathrm{pf}  \left( \frac{1}{z_i-z_j} \right)_{1\leq i \neq j \leq 2n}\]
and use \[\prod_{i=1}^{2n}z_i^{-1}\:P_{\mu_{\frac12}[n]}^{(-3)}(z_1, \ldots, z_{2n})
=P_{\mu_{\frac12}[n]-(1^{2n})}^{(-3)}(z_1, \ldots, z_{2n})
\]
where $\mu-\la=(\mu_1-\la_1,\cdots,\mu_{2n}-\la_{2n})$, so that $\mu_{\frac12}[n]-(1^{2n})$
 is the partition $\mu_{\frac12}[n]$ but with each part reduced by 1 (which is $\delta^{(2n)}(0,0)$ in the notation of \cite{BMRW2}).}
\be\label{co1}
\mathrm{pf}  \left( \frac{z_i z_j}{z_i-z_j} \right)_{1\leq i \neq j \leq 2n} =   \frac{1}{ \Delta(z_1, \ldots, z_{2n})} \, P_{\mu_{\frac12}[n]}^{(-3)}(z_1, \ldots, z_{2n})
\ee 
with
\be\mu_{\frac12}[n] = \bigl( 2n-1, 2n-1, 2n-3, 2n-3, \ldots, 1,1\bigr)\ee and
\be\label{co2}
\mathrm{pf}  \left( \frac{z_i+z_j}{z_i-z_j} \right)_{1\leq i \neq j \leq 2n} =  
\frac{1}{\Delta(z_1, \ldots, z_{2n}) } \, P_{\mu_0[n]}^{(-3)}(z_1, \ldots, z_{2n})
\ee
with 
\be \mu_0[n]=\bigl(2n-1, 2n-2,2n-3, \ldots, 1\bigr),\ee  we can then write
\be \label{exprbpta1k}\begin{split}
\Bigl[B(l;t)\, m_{\Gab}(z_1, \ldots, z_l, \ta_1, \ldots, \ta_l)\Bigr]_{\ta_1 \cdots \ta_{l-2k}} &= (-t)^k \Delta(z_1, \ldots, z_{l-2k}) \: \prod_{i=1}^l z_i^{a-l} \:\prod_{\mathclap{\substack{ 1 \leq i \leq l-2k \\  l-2k< j \leq l }}}\:(z_i-z_j) \\
&\quad \times P_{\mu_\epsilon[k]}^{(-3)}(z_{l-2k+1}, \ldots, z_l)
\end{split}
\ee
where $\epsilon=0,\tfrac12$ according to the sector of $B$.      Note that we have written the  Vandermonde determinant as
\be
\Delta(z_1,\ldots,z_{l})=\Delta(z_1, \ldots, z_{l-2k}) \, \Delta(z_{l-2k+1}, \ldots, z_{l})
\:\prod_{\mathclap{\substack{ 1 \leq i \leq l-2k \\  l-2k< j \leq l }}}\:(z_i-z_j),
\ee
to simplify the expression.  Recall that the full expression of $B(l;t)\, m_{\Gab}$ is obtained by first symmetrizing   \eqref{exprbpta1k}  and then summing  over $k$ from 0 to $\lfloor l/2\rfloor$.  \\

It follows from \cite{FJM} that the Jack polynomials evaluated at $\alpha=-3$ are well-defined for the partitions $\mu_\epsilon[k]$, (meaning that they have no poles), since these partitions satisfy the $(2,2,2k)-$admissibility condition.  Recall that a partition $\la$  is $(2,2,2k)-$admissible  if it has $2k$ parts and satisfies
\be\label{admi}\la_i-\la_{i+2}\geq 2,  \quad i=1, \ldots, 2k-2.\ee
Moreover, these Jack polynomials vanish if any three of their variables coincide (this corresponds to the so-called clustering property).    
Consequently, the right-hand side of eq.~\eqref{exprbpta1k} vanishes if any three variables in the set $\{z_{l-2k+1},\ldots,  z_l\}$ coincide.  But this is also true for three $z_i$ in the set $\{z_1, \ldots, z_{l-2k}\}$ due to the antisymmetry of the Vandermonde determinant. Moreover, the last product in the first line in  \eqref{exprbpta1k}  also vanishes whenever two variables that belong to one set coincide with one variable of the other set.   Therefore, we have (with $\Gab$ defined in \eqref{defGab})
\be
B(l;t)\, m_{\Gab}(z_1, \ldots, z_l, \ta_1, \ldots, \ta_l) = 0  \qquad \text{if} \qquad z_1=z_2=z_3.
\ee
\s

The superpolynomials that vanish when three (bosonic) variables coincide have been studied   in \cite{DLM_clust}.
Let us denote their vector space by $\mathcal{F}_l^{(2)} $, that is,
\be
\mathcal{F}_l^{(2)} = \{ f(z,\ta) \in \Sigma_l \; ; \; f(z,\ta)=0 \quad \text{if} \quad z_1=z_2=z_3 \}.
\ee
The analog of the admissibility condition \eqref{admi} for superpartitions is as follows.  
  A superpartition $\La$  is $(2,2,l)-$admissible if it satisfies
\be
\La_i^\cd -\La^*_{i+2}\geq 2, \qquad 1\leq i \leq l-2.
\ee
Denote the set of $(2,2,l)-$admissible superpartitions  by $\pi_l$ and introduce the following vector space
\be
\mathcal{I}_l^{(2,2)}:= \mathrm{span}_{\mathbb C} \{ P_\La^{(-3)} \; ; \; \La \in \pi_l \}.
\ee
It has been conjectured in \cite{DLM_clust} that
\be \mathcal{I}_l^{(2,2)}=\mathcal{F}_l^{(2)}.\ee
This conjecture implies that $B(l;t)\, m_{\Gab} \in \mathcal{I}_l^{(2,2)}$.  Quite remarkably, in the NS sector,  an explicit expression for the action of $B(l;t)$ on $m_{\Gab}$ in the sJack basis $\{P^{(-3)}_\La\}$ can be conjectured.\\

 Let $\La$ be a superpartition with exactly $m$ circles.  From left to right (equivalently, from bottom to top), mark each circle of $\La$ with $1,2,\ldots, m$. For instance
\be \La=(4,3,2,1,0;):\qquad
\tableau[scY]{&&&&\bl\tcercle{5}\\&&&\bl\tcercle{4}\\&&  \bl\tcercle{3}\\ &\bl\tcercle{2}\\\bl\tcercle{1}}.
\ee
Let $\top_i(\La)$, {for $1 \le i \le m-1$}, be the operator that acts on the diagram of $\La$ by replacing the circle marked by $i$ with
a box and removing the one marked by $i+1$. These operations may be composed, noting that we do \emph{not} relabel the circles during the intermediate operations.
For instance 
\begin{equation}
	\top_2\, (4,3,2,1,0;)=(4,3,0;2,2):\qquad
	\tableau[scY]{&&&&\bl\tcercle{5}\\&&&\bl\tcercle{4}\\&\\&\\\bl\tcercle{1}},
\end{equation}
so that
\begin{equation}
	\top_4\top_2\, (4,3,2,1,0;)=(0;4,4,2,2):\qquad
	\tableau[scY]{&&&\\&&&\\&\\&\\\bl\tcercle{1}}
\end{equation}
is defined.  Note that $\top_i$ cannot be applied twice, that $\top_{i+1}$ and $\top_i$ can never be composed, and that $\top_i \top_j = \top_j \top_i$ for all $\left\lvert i-j \right\rvert > 1$.

Given $m$, let $I=(i_1, \ldots, i_k)$, with $0\leq k\leq \lfloor m/2\rfloor$ and $1 \le i_1, \ldots, i_k \le m-1$, be an ordered set such that $i_j\geq i_{j-1}+2$.  Denote the set of all such $I$, for fixed $m$ and $k$, by $\mathcal{I}_{m,k}$.
We define the following set of superpartitions generated from {a superpartition} $\La$ with exactly $m$ circles:
\be
\mathcal{X}_k(\La) = \biggl\{\bigcup_{I {\in \mathcal{I}_{m,k}}} \,\top_{i_1}\,\top_{i_2}\cdots \top_{i_k} \,(\La) \biggr\}.
\ee
It is not difficult to show that for $k=0,1, \ldots, \lfloor l/2\rfloor$, the set  $\mathcal X_k(\Gab)$, where $\Gab = \Ga_{2a-l,l}$ is given in \eqref{defGab},  contains only $(2,2,l)$-admissible superpartitions.  
\begin{conjecture} \label{conjB1coeffPm3}
For {the} NS case, we have
\be
B(l,t) \,m_{\Gab}(z_1, \ldots, z_l, \ta_1, \ldots, \ta_l)=  
\sum_{k=0}^{\lfloor l/2\rfloor} (-t)^k 
\:\sum_{\mathclap{\La \in \mathcal{X}_k(\Gab)}}\: P_\La^{(-3)}(z_1, \ldots, z_l, \ta_1, \ldots, \ta_l).
\ee
\end{conjecture}
\s

Let us illustrate this result by considering again  the case $\Gab{= \Ga_{5,5}}
=(4,3,2,1,0;\,)$.  We first construct the sets $\mathcal{X}_k\bigl( (4,3,2,1,0;\;) \bigr)$ with $k=0,1,2$. 
We have of course $
\mathcal{X}_0\bigl( (4,3,2,1,0;\;) \bigr)=(4,3,2,1,0;\;)$. The elements of $\mathcal{X}_1$ are constructed as follows:
\begin{equation}
	\begin{aligned}
		\top_1(4,3,2,1,0;\;)&=(4,3,2;1,1), &&&&& \top_2(4,3,2,1,0;\;)&=(4,3,0;2,2), \\
		\top_3(4,3,2,1,0;\;)&=(4,1,0;3,3), &&&&& \top_4(4,3,2,1,0;\;)&=(2,1,0;4,4).
	\end{aligned}
\end{equation}
Next, for $\mathcal{X}_2$, the contributing superpartitions are
\begin{equation}
	\begin{aligned}
		\top_1\,\top_3\,(4,3,2,1,0;\;) &= (4;3,3,1,1), \\
		\top_2\,\top_4\,(4,3,2,1,0;\;) &= (0;4,4,2,2),
	\end{aligned}
	\qquad
	\top_1\,\top_4\,(4,3,2,1,0;\;) = (2;4,4,1,1).
\end{equation}
We thus finally arrive at:
\begin{equation}
	\begin{split}
		B(5;t)\, m_{(4,3,2,1,0;\,)} &=P^{(-3)}_{(4,3,2,1,0;)} -t \l(   P^{(-3)}_{(4,3,2;1,1)} + P^{(-3)}_{(4,3,0;2,2)}
		+ P^{(-3)}_{(4,1,0;3,3)}+P^{(-3)}_{(2,1,0;4,4)}\r) \\
		&\quad+ t^2 \l( P^{(-3)}_{(4;3,3,1,1)} + P^{(-3)}_{(2;4,4,1,1)}+ P^{(-3)}_{(0;4,4,2,2)} \r),
	\end{split}
\end{equation}
which {indeed} agrees with the explicit expression obtained from the action of the operator $E(5;\sqrt{t})$. 

\end{appendix}


\flushleft 

\end{document}